\documentclass[journal]{IEEEtran}
\usepackage{cite}
\usepackage{amsmath}
\usepackage{amssymb}
\usepackage{amsfonts}
\usepackage{graphicx}
\usepackage{algorithm}
\usepackage{algorithmic}
\usepackage{epsfig}
\usepackage{subfigure}
\usepackage[draft,defaultcolor=black]{changes}  
\usepackage{xcolor}
\colorlet{added}{black} 

\usepackage{pifont}
\usepackage{psfrag}
\usepackage{times}
\usepackage{booktabs}
\usepackage[T1]{fontenc}
\usepackage{moreverb}
\usepackage{color}
\usepackage{setspace}
\usepackage{mathrsfs}
\usepackage{array}
\usepackage{txfonts}
\usepackage{makecell}
\usepackage{afterpage}
\usepackage{dblfloatfix}

\begin{document}

\title{Waveform Design for Underwater Simultaneous Acoustic Information and Power Transfer}

\author{Jinheng~Kang, Yizhe~Zhao,~\IEEEmembership{Member,~IEEE}, Kun~Yang,~\IEEEmembership{Fellow,~IEEE}, Jun~Liu,~\IEEEmembership{Member,~IEEE}
\thanks{Jinheng Kang and Yizhe Zhao are with the School of Information and Communication Engineering, University of Electronic Science and Technology of China, Chengdu 611731, China,  email: jhkang@std.uestc.edu.cn, yzzhao@uestc.edu.cn.}

\thanks{Kun Yang is with State Key Laboratory of Novel Software Technology, Nanjing
	University, Nanjing 210008, China, and School of Intelligent Software and
	Engineering, Nanjing University (Suzhou Campus), Suzhou 215163, China,
	e-mail: kunyang@nju.edu.cn.}
\thanks{Jun Liu is with the School of Electronic and Information Engineering, Beihang
	University, Beijing 100191, China, e-mail: liujun2019@buaa.edu.cn.}
}

\maketitle

\begin{abstract}
Simultaneous acoustic information and power transfer (SAIPT) plays a crucial role in enabling self-sustainable and maintenance-free Internet of Underwater Things (IoUT) networks. This paper studies a multicarrier underwater SAIPT system that jointly considers the frequency-dependent characteristics of acoustic transducers and the nonlinear behavior of rectifier circuits. The waveform vector is firstly optimized using the successive convex approximation (SCA) method under constraints on average and peak transmit power for acoustic power transfer (APT). Then, in the SAIPT scenario, both the power splitting factor and waveform vectors are jointly optimized through an alternating optimization (AO) framework based on SCA, subject to transmit power and achievable rate constraints. Simulation results demonstrate that incorporating the transducer’s frequency response, rectifier nonlinearity, and the high peak-to-average power ratio (PAPR) of multicarrier waveforms leads to a significant improvement in acoustic energy transfer efficiency. The results also show that the energy harvesting DC output can be further enhanced by properly choosing system parameters, such as the number of subcarriers and subcarrier spacing.

\end{abstract}


\section{Introduction}

\subsection{Background}
	The ocean, covering approximately 70\% of the Earth's surface, plays a vital role in regulating the global climate, maintaining biodiversity, and sustaining ecosystems that support human life. It is a crucial source of food, energy, and raw materials, and serves as an indispensable medium for activities such as environmental monitoring, marine resource exploration, weather forecasting, disaster prevention, and maritime transportation \cite{linUnderwaterPollutionTracking2023}\cite{liBinocularUnderwaterMeasurement2024}. As human reliance on the ocean deepens, the demand for continuous, high-resolution, and large-scale monitoring of the marine environment has become increasingly urgent \cite{liuFundamentalsAdvancementsTopology2021}.

	To meet these needs, the Internet of Underwater Things (IoUT) has emerged as a transformative paradigm that interconnects underwater sensors, actuators, autonomous underwater vehicles (AUVs), and other intelligent nodes within a unified network. By enabling real-time data acquisition, transmission, and processing, IoUT supports a wide range of applications, including oceanographic research, underwater surveillance, pollution detection, offshore infrastructure inspection, and intelligent marine resource management\cite{chaudharyUnderwaterWirelessSensor2023}\cite{adamStateoftheArtSecuritySchemes2024}. However, the realization of IoUT faces fundamental challenges: underwater nodes are typically deployed in deep-sea or hazardous environments where battery replacement and maintenance are extremely difficult and costly, while continuous operation requires stable and reliable communication links\cite{sendraUnderwaterAcousticModems2016}.

	These constraints highlight the critical need for technologies that can simultaneously ensure information transmission and energy sustainability. Conventional underwater systems relying on battery-powered sensing nodes face severe limitations in endurance, since battery depletion leads to frequent maintenance operations and high operational costs\cite{luoSimultaneousWirelessPower2024}. To overcome these challenges, simultaneous acoustic information and power transfer (SAIPT) has emerged as a promising technique that employs acoustic waves as a unified carrier for both wireless data transmission and energy transfer. Through acoustic coupling, SAIPT enables mobile platforms such as AUVs to transmit information while simultaneously delivering energy to sensor nodes, enhancing the efficiency and sustainability of underwater networks\cite{yizhiWirelessInformationPower2024}.

	SAIPT provides a promising pathway toward self-sustaining, maintenance-free underwater networks, addressing the dual challenge of power replenishment and data connectivity. By integrating energy transfer and communication into a single acoustic framework, SAIPT lays the technological foundation for the next generation of intelligent, resilient, and energy-autonomous IoUT networks.

\subsection{Related Works}
	\added{
	Underwater communication forms the foundation of the IoUT, with acoustic communication being one of its most established and widely used approaches \cite{palCommunicationUnderwaterSensor2023}. For instance, Wang \textit{et al.} \cite{wangUnderwaterAcousticCommunications2023} proposed and validated orthogonal chirp division multiplexing (OCDM)-based techniques for improving the performance of underwater acoustic (UWA) communication under complex channel conditions. Sun \textit{et al.} \cite{sunSymbolBasedPassbandDoppler2020} proposed a symbol-by-symbol passband Doppler compensation method for DSSS UWA signals, achieving improved performance over conventional receivers. }
	
	\added{
	Multicarrier modulation has been widely adopted in underwater acoustic communication. For example, Esmaiel \textit{et al.} \cite{esmaiel2022energy} proposed a TDS-OFDM-based multicarrier system combined with NOMA to enhance spectral efficiency in underwater environments, demonstrating the effectiveness of multicarrier transmission in UWA channels. 
	Liu \textit{et al.} \cite{liu2025densenet} investigated underwater acoustic OFDM systems by developing a DenseNet-based channel estimation approach, further demonstrating the applicability of multicarrier transmission in UWA communication.	}	
	
	\added{
		These communication-oriented advances have also motivated the integration of energy transfer mechanisms into underwater acoustic systems.
	}
	In addition to data transmission, underwater wireless power transfer (WPT) has gained increasing attention as an effective solution to overcome the limited battery capacity and high maintenance cost of underwater facilities.
	Chen \textit{et al.} \cite{chenHallSensorBasedRemote2025} developed and demonstrated an underwater wireless charging system featuring a twin square compensation magnetic module, which enhances misalignment tolerance and markedly improves charging efficiency for autonomous underwater vehicles (AUVs).  The study in \cite{wibisonoSurveyUnderwaterWireless2024} provided an extensive overview of various underwater energy transfer technologies and indicated that acoustic power transfer has the advantage of high reliability in underwater environments. This is further discussed in \cite{guidaAcousticallyPoweredBatteryless2018} and \cite{guidaUnderwaterUltrasonicWireless2022}, where high-frequency acoustic waves were employed for wireless energy transfer. 
	
	Building upon the advancements in underwater communication and wireless power transfer, underwater simultaneous wireless information and power transfer (SWIPT) has emerged as a promising technology that enables concurrent energy delivery and data transmission in marine environments. There are plenty of works studying the radio-frequency (RF) based SWIPT in the mobile wireless networks. For instance, Zhang \textit{et al.}\cite{zhangEnergyEfficientPortSelection2025a} investigated a fluid-antenna-assisted IDET architecture and formulated a joint optimization of beamforming and port selection, explicitly accounting for switching delay and energy cost, to enhance both short-term and long-term wireless energy transfer efficiency. Bian \textit{et al.}\cite{bianQoSAwareEnergyStorage2023} addressed reconfigurable intelligent surface (RIS)-assisted wireless networks by jointly optimizing SWIPT and mobile edge computing (MEC), where RIS coefficients and task allocation were coordinated via a block coordinate descent (BCD)-based algorithm to maximize energy storage under quality-of-service (QoS) constraints. Huang \textit{et al.}\cite{huangHolographicIntegratedData2024} studied holographic integrated data and energy transfer (H-IDET) and developed a BCD-based beamforming design to maximize data-user sum-rate while meeting energy-user harvesting requirements, leveraging the electromagnetic manipulation capability of holographic MIMO. Li \textit{et al.} \cite{liDynamicUserSchedulingPower2023} formulated energy-aware federated learning with SWIPT and non-orthogonal multiple access (NOMA), and developed a proximal-policy-optimization-based actor-critic algorithm to jointly optimize device scheduling and power control for minimizing long-term energy consumption. Zhao \textit{et al.}\cite{zhaoReflectiveIndexModulation2024} proposed an IRS-assisted IDET scheme using reflective index modulation (RIM) to jointly enhance wireless data transfer and energy harvesting, and optimized IRS phase shifters to maximize harvested energy under data-rate and BER constraints. In underwater scenarios, RF-based SWIPT is no longer applicable, due to the significant power attenuation of the transmission of electromagnetic waves. Ye \textit{et al.}\cite{yeDualHopUnderwaterOptical2021} proposed a dual-hop underwater optical wireless communication (UOWC) system employing simultaneous lightwave information and power transfer (SLIPT) to extend communication range and significantly enhance system performance. Guo \textit{et al.} \cite{guoJointDesignCommunication2021} developed a magnetic induction-based mobile underwater charging solution that enables SWIPT for AUV swarms, effectively extending their operating range and mission duration. \added{For acoustic-based SWIPT (SAIPT), Xing \textit{et al.} \cite{xing2023performance} provided closed-form performance analysis for a SAIPT-enabled system, with explicit characterization of BER and outage behavior under different signal frequency and time allocation strategies. Esmaiel \textit{et al.} \cite{esmaiel2020wireless} investigated a time-reversed NOMA (TR-NOMA) for SAIPT, demonstrating performance gains over conventional NOMA by leveraging passive time-reversal. Omeke \textit{et al.} \cite{omekeSustainableInternetUnderwater2024a} proposed a reinforcement learning-based underwater SAIPT scheme using an AUV to simultaneously optimize data throughput and wireless power transfer, significantly improving the overall energy efficiency of underwater networks. Deepa \textit{et al.} \cite{deepaEnhancedSWIPTCooperative2025} presented an enhanced SAIPT scheme with cooperative relaying for NOMA-based underwater acoustic sensor networks, achieving improved data rate, energy efficiency, and communication reliability. Melki \textit{et al.} \cite{melkiAUVTrajectoryLearning2025} introduced a deep reinforcement learning (DRL)-based acoustic SAIPT framework using an AUV for sustainable IoUT operations, integrating the age of information (AoI) concept to jointly enhance data freshness, energy harvesting, and fairness. }

\subsection{Motivations and Contributions}	
	\added{SAIPT systems have recently gained growing attention for enabling sustainable IoUT networks. Despite the advances, existing studies often rely on simplified device-level modeling, which limits their practical applicability. For instance, \cite{xing2023performance} assumes a frequency-independent transducer conversion efficiency, which does not capture the practical frequency-selective characteristics of acoustic transducers. In addition, \cite{esmaiel2020wireless} investigates a SAIPT scheme where, although the diode nonlinearity is initially modeled using a Taylor series expansion, the harvested DC output is ultimately approximated using only the second-order term, thereby neglecting higher-order nonlinear effects.} These limitations lead to inaccurate performance evaluation and hinder the effective design of waveform optimization strategies in practical underwater environments.
	
	These modeling limitations motivate a deeper examination of the trade-offs in SAIPT waveform design, 
	which arise from multiple conflicting objectives:
	\begin{itemize}
		\item Due to the limited high-efficiency electro-acoustic and acousto-electric conversion bandwidth 
		of the transducer, the subcarriers should be concentrated around its resonance frequency, with a 
		preference for a smaller number of subcarriers and narrower subcarrier spacing.
		\item Owing to the nonlinear characteristics of the rectification circuit in the energy-harvesting 
		stage, increasing the number of subcarriers boosts the peak-to-average power ratio (PAPR) of the 
		transmit signal, which in turn enhances the wireless power transfer efficiency.
		\item Enlarging the subcarrier spacing expands the overall signal bandwidth, reducing the power 
		required for information transmission and allowing more transmit power to be allocated for power 
		transfer.
	\end{itemize}
		
	Building on these insights, this work develops a comprehensive SAIPT framework for waveform design that jointly considers the frequency-dependent characteristics of acoustic transducers and the nonlinear energy conversion behavior of rectifiers. The main contributions of this paper are summarized as follows:
	\begin{itemize}
	\item A multicarrier underwater acoustic-based transmission model is established, where the receiver adopts a power-splitting architecture for simultaneous information and power reception, and a nonlinear rectifier model is introduced to capture the diode-based energy conversion behavior.
	
	\item In the underwater acoustic power transfer (APT) case, the waveform vector is optimized using the successive convex approximation (SCA) method to maximize the transferred power under the average transmit power and peak power constraints.
	
	\item In the SAIPT scenario, the power splitting factor and waveform vectors are jointly optimized through an alternating optimization (AO) framework based on SCA under the constraints of average transmit power, peak power, and achievable information rate.
	
	\item Simulation results verify that incorporating transducer characteristics and the nonlinear rectification behavior, together with exploiting the high PAPR of multicarrier waveforms, can significantly improve power transfer efficiency. In addition, the results show that appropriately enlarging the subcarrier spacing reduces the energy required for information transmission, thereby allowing more transmit power to be allocated to power transfer. Overall, the findings highlight the importance of proper system parameter selection for achieving optimal performance.
	
	\end{itemize}
	
	The rest of this paper is organized as follows: The model of SAIPT system is introduced in Section \ref{sec:SYSTEM MODEL}. The APT optimization and SAIPT optimization for maximizing the DC output are presented in Section \ref{sec:WAVEFORM DESIGN FOR APT} and Section \ref{sec:WAVEFORM DESIGN FOR SAIPT}, respectively. Simulation results are illustrated in Section \ref{sec:SIMULATION RESULTS}. Finally, our paper is concluded in Section \ref{sec:CONCLUSION}.
	
	\textit{Notations:} In this paper, boldface upper letters, boldface lower letters, and lower letters denote matrices, vectors, and scalars, respectively. 
	$\mathbb{C}^{m \times n}$ and $\mathbb{R}^{m \times n}$ represent the set of complex matrices and real matrices with the dimension of $m \times n$, respectively. 
	A complex vector of dimension $m$ is denoted in $\mathbb{C}^{m}$, while a real vector of dimension $m$ is denoted in $\mathbb{R}^{m}$. 
	$|x|$ represents the amplitude of a complex number $x$. 
	$\|\mathbf{x}\|$  denotes the 2-norm of a vector $\mathbf{x}$. 
	The notation $(\cdot)^{T}$ and $(\cdot)^{H}$ refer to the transpose and the conjugate transpose of a vector or matrix. $\mathbb{E}$ represents the expectation operator, while $\mathcal{P}\{\cdot\}$ refers to the DC component of a signal. $\Re\{\cdot\}$ and $\Im\{\cdot\}$ denote the real and imaginary parts of a complex vector or matrix, respectively.

\section{SYSTEM MODEL} \label{sec:SYSTEM MODEL}
	\begin{figure*}[htbp]
	\centering
	\includegraphics[width=0.8\textwidth]{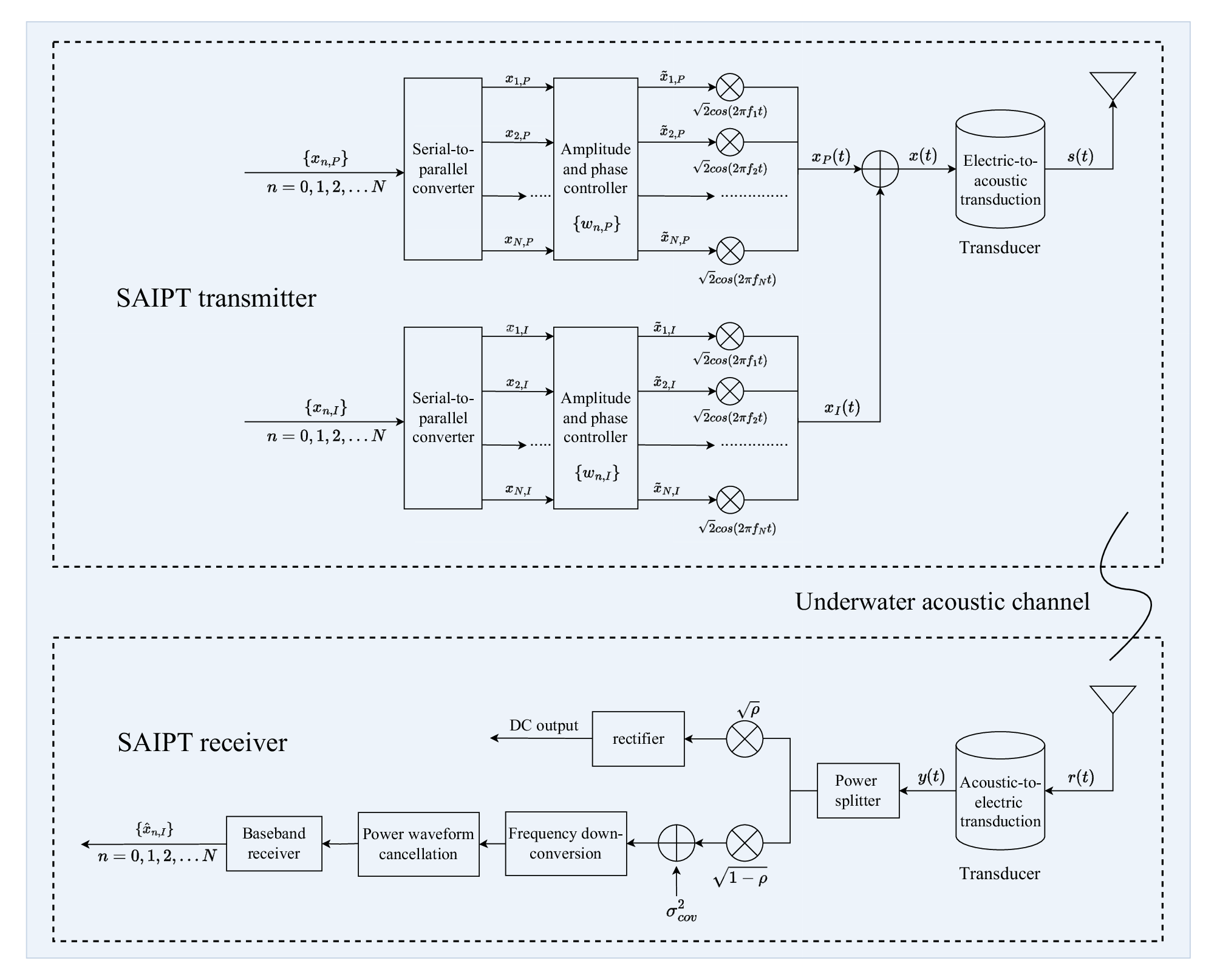}
	\caption{SAIPT system model.}
	\label{fig:system_flowchart}
	\end{figure*}
\subsection{SAIPT Model} \label{subsec:SAIPT Model}

An OFDM-based SAIPT system is illustrated in Fig.~\ref{fig:system_flowchart}, which consists of a SAIPT transmitter and a SAIPT receiver. We consider an OFDM system with $N$ orthogonal subcarriers, where the $n$-th subcarrier has frequency $f_n$ and an identical subcarrier spacing $\Delta f$, with $n = 1, 2, \ldots, N$. The carrier frequencies are uniformly spaced and can be expressed as
\begin{equation}
	f_n = f_r + \left(n - \frac{N+1}{2}\right) \Delta f,
\end{equation}
where $f_r$ denotes the resonant frequency of the transducer.

The SAIPT transmitter is illustrated at the top of Fig.~\ref{fig:system_flowchart}. Specifically, $x_{n,P} $ denotes the deterministic power-transfer symbol (set to unity for all subcarriers, i.e., $x_{n,P}=1, ~ \forall n$) and $x_{n,I} \sim \mathcal{CN}(0,1)$ denotes the information-transfer symbol on the $n$-th subcarrier. Collecting all subcarriers, the symbol vectors are defined as
$
\mathbf{x}_{P} = [x_{1,P}, x_{2,P}, \ldots, x_{N,P}]^{T} \in \mathbb{C}^{N}, ~
\mathbf{x}_{I} = [x_{1,I}, x_{2,I}, \ldots, x_{N,I}]^{T} \in \mathbb{C}^{N}.
$

For both power transfer and information transfer, the amplitude--phase controller adjusts
each element of $\mathbf{x}_{P}$ and $\mathbf{x}_{I}$ using the corresponding coefficient 
vectors $\mathbf{w}_{P} = [w_{1,P},\ldots,w_{N,P}]^{T} \in \mathbb{C}^{N}$ and 
$\mathbf{w}_{I} = [w_{1,I},\ldots,w_{N,I}]^{T} \in \mathbb{C}^{N}$, respectively. 
The adjusted vectors are given by
$
\tilde{\mathbf{x}}_{P} = \mathbf{w}_{P} \odot \mathbf{x}_{P},~
\tilde{\mathbf{x}}_{I} = \mathbf{w}_{I} \odot \mathbf{x}_{I},
$
where $\odot$ denotes the element-wise product. 
Each modified vector is then modulated onto its corresponding subcarriers and 
superimposed to form the electrical signals $x_{P}(t)$ and $x_{I}(t)$ for power 
and information transfer, respectively.
The integrated transmitted signal $x(t)$ is obtained by superimposing the power- and information-transfer signals. After passing through the transducer, the integrated transmitted signal $x(t)$ is converted into the SAIPT signal $s(t)$ and transmitted into the underwater acoustic channel.

After propagation over a distance of $d$ m, the received SAIPT signal $r(t)$ is first converted into the integrated received signal $y(t)$ by the receiver transducer. Using the power splitter with splitting factor $\rho$, the signals $\sqrt{\rho}\,y(t)$ and $\sqrt{1-\rho}\,y(t)$ are then respectively directed to the rectifier for DC output and to the information decoder.\footnote{\added{In this work, we focus on the end-to-end signal processing from $x(t)$ to $y(t)$. 
The electro-acoustic-electrical (EAE) conversion process is incorporated into the system model through the transducer conversion efficiency. Therefore, the intermediate conversions from $x(t)$ to $s(t)$ and from $r(t)$ to $y(t)$ are not explicitly modeled in detail.}}

\subsection{Underwater Channel Model}
	
	\subsubsection{\added{Large-Scale Dominant Channel Coefficient}}
	\label{sec:channel coefficient}
	\added{We consider a SAIPT scenario with a relatively short transmission distance, where the underwater acoustic link is dominated by the direct propagation path, while the contribution of small-scale multipath components is relatively weak. As a result, the channel coefficient is primarily determined by large-scale effects, and small-scale fading is not explicitly modeled.}
	
	\added{The large-scale path loss (in dB) is modeled as a function of the transmission distance $d$ and signal frequency $f$ \cite{stojanovicRelationshipCapacityDistance2007}:
	\begin{equation}
			PL(f,d) = k \, 10 \log(d) + \alpha(f)d ,
	\end{equation}
	where $k$ denotes the spreading factor and $\alpha(f)$ is the frequency-dependent absorption coefficient.}
	
	\added{Based on the above path loss model, the underwater acoustic channel coefficient at the $n$-th subcarrier is expressed as
	\begin{equation}
		h_n = \hat{h}_n + \Delta h_n,
	\end{equation}
	where $\hat{h}_n$ denotes the estimated channel coefficient at frequency $f_n$ obtained from the large-scale path loss model, i.e.,
	\begin{equation}
		\hat{h}_n=\frac{1}{10^{PL(f_n,d)/20}}.
	\end{equation}
	$\Delta h_n \sim \mathcal{N}(0,\sigma_e^2 |\hat{h}_n|^2) $ represents the channel estimation error accounting for residual uncertainty arising from weak multipath components and temporal variations, and $\sigma_e^2$ denotes the magnitude of the relative channel estimation error.}

	\subsubsection{\added{Underwater Noise}}
	\added{
	Underwater acoustic noise consists of four main components: turbulence, shipping, wind, and thermal noise. 
	The overall power spectral density (p.s.d.) is modeled as the sum of these components:}
	\added{
	\begin{equation}
		N(f)=N_{t}(f)+N_{s}(f)+N_{w}(f)+N_{th}(f).
	\end{equation}
	Detailed expressions of each noise component can be found in \cite{basagniAdvancesUnderwaterAcoustic2013}.}
	
	\added{The noise power over the system bandwidth $B$ is then obtained by integrating the p.s.d. as \cite{wangOptimalPowerAllocation2020a}
	\begin{align}
		P_{n}&=10\log_{10}\left(\int_{f_0}^{f_0+B}N(f)df\right),\\
		\sigma_{n}^{2}&=10^{\frac{P_{n}-10\log_{10}\phi-171.5}{10}},
	\end{align}
	where $f_{0}$ denotes the minimum operational frequency, $\phi$ represents the overall electro-acoustic efficiency, and $\sigma_n^2$ denotes the underwater noise power in watts.}

	\subsection{\added{Transducer Conversion Efficiency}} \label{sec:eta_n}
	\added{We define the electro-acoustic-electrical (EAE) conversion efficiency at frequency $f$ as 
	\begin{equation}
		\eta(f) = \frac{P_{\text{harv}}(f)}{P_{\text{elec}}(f)},
	\end{equation}
	where $P_{\text{elec}}(f)$ and $P_{\text{harv}}(f)$ denote the transmitted and harvested electrical power at frequency $f$, respectively.
	This efficiency characterizes the overall EAE energy conversion performed by the transmit and receive transducers, while excluding the propagation loss.}
	
	\added{According to the analytical results reported in \cite{akbarUnderwaterBackscatterChannel2023}, the relationship between $P_{\text{harv}}(f)$ and $P_{\text{elec}}(f)$ is given by
	\begin{equation}
		\begin{aligned}
			P_{\text{harv}}(f)
			&=47.7+10 \log \left(P_{\text{elec}}(f)\right)
			+G_{\text{TX}}(f)\\
			&\quad+G_{\text{node}}(f)-20 \log (f),
			\label{eq:P_harv_dB_EAE}
		\end{aligned}
	\end{equation}
	where $P_{\text{elec}}(f)$ is measured in watts, $P_{\text{harv}}(f)$ is in dBW, and
	$G_{\text{TX}}(f)$ and $G_{\text{node}}(f)$ denote the transmitter and receiver transducer gains (in dB) at frequency $f$, respectively.}
	
	\added{Transforming \eqref{eq:P_harv_dB_EAE} into the linear domain yields
	\begin{equation}
		P_{\text{harv}}(f)
		=\frac{10^{4.77}\, P_{\text{elec}}(f)\,
			G_{\text{TX}}^{\prime}(f)\,
			G_{\text{node}}^{\prime}(f)}{f^{2}},
		\label{eq:P_harv_linear_EAE}
	\end{equation}
	where $G_{\text{TX}}^{\prime}(f)$ and $G_{\text{node}}^{\prime}(f)$ denote the corresponding linear transducer gains.
	The explicit expressions of the transducer gains will be provided in Section~\ref{sec:Transducer Gain Fitting}.}
	
	\added{In practical system design, if identical transducers are employed at the transmitter and receiver, i.e., $G_{\text{TX}}^{\prime}(f)=G_{\text{node}}^{\prime}(f)=G^{\prime}(f)$, the harvested power can be simplified as
	\begin{equation}
		P_{\text{harv}}(f)
		=\frac{10^{4.77}\, P_{\text{elec}}(f)\, G^{\prime 2}(f)}{f^{2}}.
	\end{equation}}
	
	\added{Accordingly, the EAE conversion efficiency is given by
	\begin{equation}
		\eta(f)
		=\frac{P_{\text{harv}}(f)}{P_{\text{elec}}(f)}\Big|_{P_{\text{elec}}(f)=1\text{ W}}
		=\frac{10^{4.77}\, G^{\prime 2}(f)}{f^{2}}.
	\end{equation}}
	
	\added{Furthermore, $\eta_n$ is defined as the EAE conversion efficiency of the $n$-th subcarrier, which can be expressed as $\eta_n = \eta(f_n).$}

\subsection{Signal Model}
	As stated in Section~\ref{subsec:SAIPT Model}, the integrated transmitted signal $x(t)$ can be expressed as
	\begin{equation}
		x(t)
		= \sqrt{2}\Re\left\{ \sum_{n=1}^{N} \big( w_{n,I}x_{n,I} + w_{n,P}x_{n,P} \big) e^{j2\pi f_n t} \right\} \triangleq x_{I}(t) + x_{P}(t),
	\end{equation}
	where  $\displaystyle x_{I}(t)
	\triangleq \sqrt{2}\Re\!\left\{ \sum_{n=1}^{N} w_{n,I} x_{n,I} e^{j2\pi f_n t} \right\}$ and $ \displaystyle x_{P}(t)
	\triangleq \sqrt{2}\Re\!\left\{ \sum_{n=1}^{N} w_{n,P} x_{n,P} e^{j2\pi f_n t} \right\}$.
	\vspace{1em}
	
	According to Sections~\ref{sec:channel coefficient} and~\ref{sec:eta_n}, the integrated received signal $y(t)$ can be expressed as
	\added{
	\begin{equation}
		\begin{aligned}
		y(t) &= \sqrt{2}\Re\left\{ \sum_{n=1}^{N} \sqrt{\eta_{n}}h_n \big(w_{n,I}x_{n,I} + w_{n,P}x_{n,P}\big) e^{j2\pi f_n t} \right\}\\
		&\triangleq y_{I}(t) + y_{P}(t),
		\end{aligned}
	\end{equation}
	where $y_{I}(t)$ and $y_{P}(t)$ denote the received electrical information-transfer and power-transfer signals, respectively, which are defined as 
	\begin{align}
		y_{I}(t) &\triangleq \sqrt{2}\Re\left\{\sum_{n=1}^{N} \sqrt{\eta_{n}}h_n w_{n,I} x_{n,I} e^{j2\pi f_{n}t}\right\}, \\
		y_{P}(t) &\triangleq \sqrt{2}\Re\left\{\sum_{n=1}^{N} \sqrt{\eta_{n}}h_n w_{n,P} x_{n,P} e^{j2\pi f_{n}t}\right\}.
	\end{align}}
	
	\subsection{Performance Analysis}
	\subsubsection{Acoustic Power Transfer Performance}
	In the energy-harvesting branch, the power-split signal $y_{in}(t)=\sqrt{\rho}y(t)$ is rectified to yield a DC output, which is subsequently used to power the underwater node.
	
	According to the nonlinear diode model of the rectifier, the relationship between the DC output and the input signal $y_{\mathrm{in}}(t)$ can be expressed as \cite{clerckxWaveformDesignWireless2016}
	
	\begin{equation}
		z_{\mathrm{DC}}
		= k_{2} R_{\text{trans}}\, \mathbb{E}\!\left[ \mathcal{P}\!\left( y_{\mathrm{in}}^{2}(t) \right) \right]
		+ k_{4} R_{\text{trans}}^{2}\, \mathbb{E}\!\left[ \mathcal{P}\!\left( y_{\mathrm{in}}^{4}(t) \right) \right],
		\label{eq:zDC_origin}
	\end{equation}
	where $k_{2}$ and $k_{4}$ denote the constants of the rectifier, $R_{\text{trans}}$ represents the impedance of the transducer, and $z_{\mathrm{DC}}$ is proportional to the DC output power. Therefore, maximizing the DC output power is equivalent to maximizing $z_{\mathrm{DC}}$.
	
	Given that $\mathbb{E}\!\left[\mathcal{P}(y_{P}(t)y_{I}(t))\right]= 
	\mathbb{E}\!\left[\mathcal{P}(y_{P}(t)^{3}y_{I}(t))\right]= 
	\mathbb{E}\!\left[\mathcal{P}(y_{P}(t)y_{I}(t)^{3})\right]=0$, 
	and $y_{P}(t)$ is a deterministic signal, \eqref{eq:zDC_origin} can be reformulated as 
	\begin{align}
		z_{\mathrm{DC}} = &\; k_{2}R_{\text{trans}}\rho\,
		\mathbb{E}\!\left[\mathcal{P}(y_{P}^{2}(t))\right]
		+ k_{4}R_{\text{trans}}^{2}\rho^{2}\,
		\mathbb{E}\!\left[\mathcal{P}(y_{P}^{4}(t))\right] \notag \\[2pt]
		&+ k_{2}R_{\text{trans}}\rho\,
		\mathbb{E}\!\left[\mathcal{P}(y_{I}^{2}(t))\right]
		+ 6k_{4}R_{\text{trans}}^{2}\rho^{2}\,
		\mathbb{E}\!\left[\mathcal{P}(y_{P}^{2}(t))\right]
		\mathbb{E}\!\left[\mathcal{P}(y_{I}^{2}(t))\right] \notag \\[2pt]
		&+ k_{4}R_{\text{trans}}^{2}\rho^{2}\,
		\mathbb{E}\!\left[\mathcal{P}(y_{I}^{4}(t))\right].
		\label{eq:zDC_2nd}
	\end{align}
	
	\added{
	Furthermore, define the estimated effective channel vector for all the subcarriers as
	\begin{equation}
		\hat{\mathbf{h}}_{\mathrm{eff}}
		= \left[ \sqrt{\eta_1} \hat{h}_1, \sqrt{\eta_2} \hat{h}_2, \cdots, \sqrt{\eta_N} \hat{h}_N \right]^T \in \mathbb{R}^{N},
	\end{equation}
	and the matrix as
	\begin{equation}
		\mathbf{M} = \hat{\mathbf{h}}_{\mathrm{eff}}\hat{\mathbf{h}}_{\mathrm{eff}}^T.
	\end{equation}}
	
	\added{To facilitate subsequent analysis, we further introduce a family of auxiliary matrices
	$\mathbf{M}_k$. Specifically, $\mathbf{M}_k$ is an $N\times N$ matrix obtained by
	preserving only a selected subset of the entries of $\mathbf{M}$, with all other
	entries set to zero. For $k \in \{1,\ldots,N-1\}$, the matrix $\mathbf{M}_k$ retains
	the elements located on the $k$-th super-diagonal of $\mathbf{M}$; and
	for $k=0$, it preserves the main diagonal of $\mathbf{M}$.}
	
	\added{Then each term in \eqref{eq:zDC_2nd} can be rewritten in the matrix form as}
	\begingroup 
	\definecolor{added}{RGB}{0,0,0} 
	\color{added}                     
	\everymath{\color{added}}         
	
	\begin{IEEEeqnarray}{rCl}
		\color{added}
		\mathbb{E}\!\left[\mathcal{P}\big(y_{P}^{2}(t)\big) \right]
		&=& \sum_{n=1}^{N}\eta_{n}\hat{h}_n^2\lvert w_{n,P}\rvert^{2} = \mathbf{w}_{P}^{H}\mathbf{M}_{0}\mathbf{w}_{P}, \label{eq:matrix1}\\    
		\mathbb{E}\!\left[\mathcal{P}\big(y_{P}^{4}(t)\big) \right]
		&=&\frac{3}{2}\sum_{\substack{n,n',m,m'\\ f_n+f_{n'}=f_m+f_{m'}}}
		\Big(\sqrt{\eta_{n}\eta_{n'}\eta_{m}\eta_{m'}}\,
		\IEEEnonumber\\
		& \,&  \hat{h}_{n}\hat{h}_{n'}\hat{h}_{m}\hat{h}_{m'}w_{n,P}w_{n',P}w_{m,P}^{*}w_{m',P}^{*}\Big) \IEEEnonumber\\
		&=& \frac{3}{2}\big(\mathbf{w}_{P}^{H}\mathbf{M}_{0}\mathbf{w}_{P}\big)^{2}
		+3\sum_{k=1}^{N-1}\big(\mathbf{w}_{P}^{H}\mathbf{M}_{k}\mathbf{w}_{P}\big)^{2}, \\
		\mathbb{E}\big[\mathcal{P}(y_{I}^{2}(t))\big]
		&=& \sum_{n=1}^{N} \eta_{n} \hat{h}_n^2\,\lvert w_{n,I}\rvert^{2} 
		=\mathbf{w}_{I}^{H}\mathbf{M}_{0}\mathbf{w}_{I},\\
		\mathbb{E}\big[\mathcal{P}(y_{I}^{4}(t))\big]
		&=& 3\left[ \sum_{n=1}^{N}
		\eta_{n} \hat{h}_n^2\,\lvert w_{n,I}\rvert^{2} \right]^{2}= 3\left(\mathbf{w}_{I} 
		^{H}\mathbf{M}_{0}\mathbf{w}_{I} \right)^{2}. \label{eq:matrix_end}
	\end{IEEEeqnarray}
	\endgroup
	
	\added{
	When taking expectation, the zero-mean channel estimation error terms vanish, and hence the matrices $\mathbf{M}_k$ in \eqref{eq:matrix1}--\eqref{eq:matrix_end} are constructed from the estimated large-scale coefficients.}

	\added{
	By substituting \eqref{eq:matrix1} -- \eqref{eq:matrix_end} into \eqref{eq:zDC_2nd}, the  $z_{\mathrm{DC}}$ expression can be rewritten in the matrix form as
	\begin{equation}
		\label{eq:zDC_matrix}
		\begin{aligned}
			z_{\mathrm{DC}}
			&= k_{2}R_{\text{trans}}\rho\mathbf{w}_{P}^{H}\mathbf{M}_{0}\mathbf{w}_{P}  + \tfrac{3}{2}k_{4}R_{\text{trans}}^{2}\rho^{2}
			\left(\mathbf{w}_{P}^{H}\mathbf{M}_{0}\mathbf{w}_{P}\right)^{2}\\
			& + 3k_{4}R_{\text{trans}}^{2}\rho^{2}
			\sum_{k=1}^{N-1}\left(\mathbf{w}_{P}^{H}\mathbf{M}_{k}\mathbf{w}_{P}\right)^{2} 
			+ k_{2}R_{\text{trans}}\rho \mathbf{w}_{I}^{H}\mathbf{M}_{0}\mathbf{w}_{I} \\
			& + 6k_{4}R_{\text{trans}}^{2}\rho^{2}
			\left(\mathbf{w}_{P}^{H}\mathbf{M}_{0}\mathbf{w}_{P}\right)
			\left(\mathbf{w}_{I}^{H}\mathbf{M}_{0}\mathbf{w}_{I}\right)  \\
			&+ 3k_{4}R_{\text{trans}}^{2}\rho^{2}
			\left(\mathbf{w}_{I}^{H}\mathbf{M}_{0}\mathbf{w}_{I}\right)^{2}.
		\end{aligned}
	\end{equation}}
	\added{
	Furthermore, $\mathbf{w}_P,\mathbf{w}_I$ and $\mathbf{M}_k$ can be decomposed into their real and imaginary parts, thereby converting the expressions from the complex domain to the real domain and reducing the complexity of subsequent optimization. Accordingly, the expression of $z_{DC}$ can be reformulated as
	\begin{equation}
		\label{eq:zDC_matrix_real}
		\begin{aligned}
		z_{\mathrm{DC}} 
		&= k_{2}R_{\text{trans}}\rho \mathbf{\hat{w}}_{P}^{T}\mathbf{N}_{0}\mathbf{\hat{w}}_{P}
		+ \frac{3}{2}k_{4}R_{\text{trans}}^{2}\rho^{2}\left(\mathbf{\hat{w}}_{P}^{T}\mathbf{N}_{0}\mathbf{\hat{w}}_{P}\right)^{2}\\
		&+ 3k_{4}R_{\text{trans}}^{2}\rho^{2}\sum_{k=1}^{N-1}\left(\mathbf{\hat{w}}_{P}^{T}\mathbf{N}_{k}\mathbf{\hat{w}}_{P}\right)^{2} 
		 + k_{2}R_{\text{trans}}\rho \mathbf{\hat{w}}_{I}^{T}\mathbf{N}_{0}\mathbf{\hat{w}}_{I} \\
		&+ 6k_{4}R_{\text{trans}}^{2}\rho^{2}\left(\mathbf{\hat{w}}_{P}^{T}\mathbf{N}_{0}\mathbf{\hat{w}}_{P}\right)
		\left(\mathbf{\hat{w}}_{I}^{T}\mathbf{N}_{0}\mathbf{\hat{w}}_{I}\right)\\
		&+ 3k_{4}R_{\text{trans}}^{2}\rho^{2}\left(\mathbf{\hat{w}}_{I}^{T}\mathbf{N}_{0}\mathbf{\hat{w}}_{I}\right)^{2},
		\end{aligned}
	\end{equation}
	where 
	$\mathbf{\hat{w}}_{P} = \big(\Re\{\mathbf{w}_{P}\}, \Im\{\mathbf{w}_{P}\}\big)^{T}\triangleq
	 [\hat{w}_{1,P}, \hat{w}_{2,P}, \ldots, \hat{w}_{2N,P}]^{T} \in \mathbb{R}^{2N}$, 
	$\mathbf{\hat{w}}_{I} = \big(\Re\{\mathbf{w}_{I}\}, \Im\{\mathbf{w}_{I}\}\big)^{T}\triangleq
	[\hat{w}_{1,I}, \hat{w}_{2,I}, \ldots, \hat{w}_{2N,I}]^{T} \in \mathbb{R}^{2N}$, 
	and
	$
	\mathbf{N}_{k}=
	\begin{pmatrix}
		\Re\{\mathbf{M}_{k}\} & -\Im\{\mathbf{M}_{k}\} \\
		\Im\{\mathbf{M}_{k}\} & \Re\{\mathbf{M}_{k}\}
	\end{pmatrix}\in \mathbb{R}^{2N \times 2N}.
	$}
	\vspace{1em}
	
	\subsubsection{Acoustic Information Transfer Performance }
	\added{After passing through the power splitter, the signal component 
	$\sqrt{1-\rho}\,y(t)$ is allocated to information transfer. 
	According to Shannon's capacity formula, the achievable data rate of the system 
	can be expressed as  
	\begin{equation}
	\label{eq:R_I}
	R_{I}  = \sum_{n=1}^{N}\Delta f \log_{2}\left(1+\frac{(1-\rho)\eta_{n}(1+\sigma_{e}^{2})\hat{h}_{n}^{2}|w_{n,I}|^{2}}{(1-\rho)\sigma_{n}^{2}+\sigma_{cov}^{2}+(1-\rho)\eta_{n}\hat{h}_{n}^{2}\sigma_{e}^{2}|w_{n,P}|^{2}}\right)
	\end{equation}
	where $\sigma_n^{2}$ and $\sigma_{\text{cov}}^{2}$ represent the noise  power of the underwater acoustic channel at the $n$-th subcarrier and the noise power induced by the down-conversion process, respectively.}
	
\subsection{Transducer Gain Fitting} \label{sec:Transducer Gain Fitting}
	The transducer gain can be calculated from its transmitting voltage response (TVR) and impedance frequency response \cite{akbarUnderwaterBackscatterChannel2023} as
	\begin{equation}
		G(f) = TVR(f) - 10 \log \!\left(\frac{\Re\{Z(f)\}}{|Z(f)|^2}\right) - 170.8,
		\label{gain_dB}
	\end{equation}
	where $TVR(f)$ denotes the relationship between the voltage excitation and the 
	acoustic pressure response of the transducer at frequency $f$ (in dB re 1~$\mu$Pa/V @ 1m), 
	and $Z(f)$ represents the complex impedance of the transducer at the same frequency.
	The linear transducer gain $G'(f)$ is related to $G(f)$ through
	$G'(f) = 10^{\frac{G(f)}{10}}.$
	
	According to the experimental data in \cite{allamDemoUnderwaterBackscatter2023}, the measured TVR and impedance of the BII-7511 transducer can be used with \eqref{gain_dB} to obtain discrete linear gain values through linear transformation. Since these measurements are available only at limited sampling points, curve fitting is required to derive a continuous gain function over the entire operating band for system modeling and waveform optimization.
	
	Several curve fitting methods were applied to the BII-7511 transducer data \cite{allamDemoUnderwaterBackscatter2023} using the MATLAB Curve Fitting Toolbox. The fitting results are summarized in Table~\ref{tab:BII-7511_curve_fit}.
	
	\begin{table*}[!t]
		\centering
		\caption{Curve fitting results for the BII-7511 transducer gain}
		\label{tab:BII-7511_curve_fit}
		\begin{tabular}{lccccc}
			\toprule
			\textbf{Fitting method} &
			\textbf{$R^2$} &
			\makecell{\textbf{Sum of}\\\textbf{Squared Errors}} &
			\makecell{\textbf{Adjusted}\\\textbf{$R^2$}} &
			\makecell{\textbf{Root Mean}\\\textbf{Squared Error}} &
			\makecell{\textbf{Number of}\\\textbf{coefficients}} \\
			\midrule
			Sinusoidal sum & 0.99744 & 0.054108 & 0.99729 & 0.011996 & 24 \\
			Fourier        & 0.99584 & 0.087980 & 0.99566 & 0.015176 & 18 \\
			Gaussian       & 0.99515 & 0.102650 & 0.99490 & 0.016457 & 21 \\
			Polynomial     & 0.98598 & 0.296710 & 0.98566 & 0.027582 & 10 \\
			Sigmoidal      & 0.82874 & 3.625600 & 0.82744 & 0.095685 & 4  \\
			Power          & 0.65215 & 7.364000 & 0.65040 & 0.136200 & 3  \\
			Rational       & 0.65020 & 7.405300 & 0.64121 & 0.137970 & 11 \\
			Logarithmic    & 0.62007 & 8.043100 & 0.61912 & 0.142160 & 2  \\
			\bottomrule
		\end{tabular}
	\end{table*}

\begin{figure}[htbp]   
		\centering         
		\includegraphics[width=0.8\columnwidth]{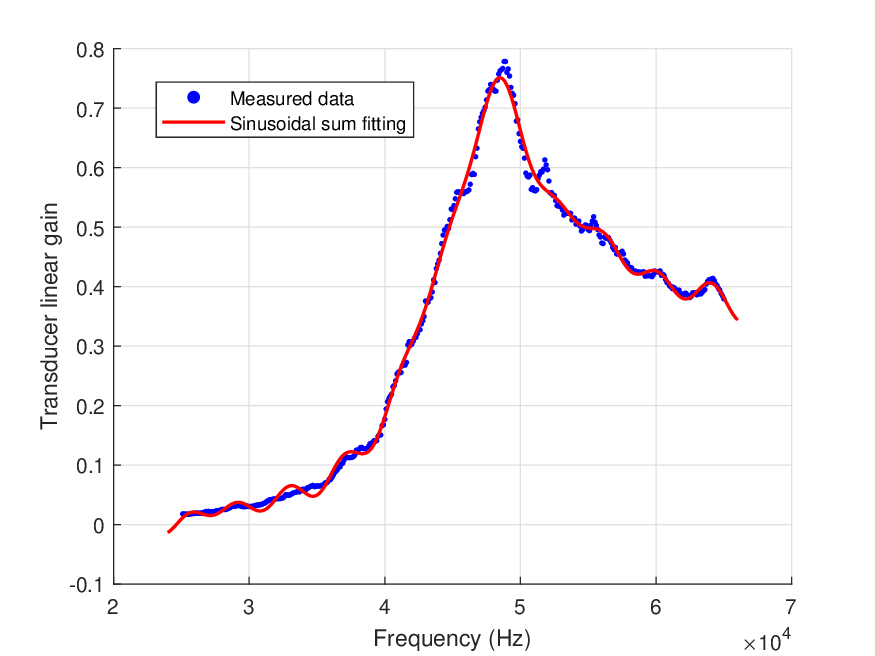} 
		\setlength{\abovecaptionskip}{0pt}
		\caption{Measured and fitted gain of the BII-7511 transducer.} 
		\label{fig:BII-7511-fitting} 
	\end{figure}

	The sinusoidal sum model achieved the best performance for the BII-7511 transducer, accurately capturing its resonant gain characteristics. Despite involving 24 parameters, its smoothness and fitting accuracy make it the most suitable choice for subsequent system modeling, as illustrated in Fig.~\ref{fig:BII-7511-fitting}.
	
	The sinusoidal sum model is expressed as  
	$	
		G_{\text{fit}}(f) = \sum_{i=1}^{8} a_i \cdot \sin(b_i f + c_i),
	$
	where $a_i$, $b_i$, and $c_i$ denote the amplitude coefficient, frequency scaling factor, and phase offset of the $i$th sinusoidal basis function, respectively, all of which are obtained from the gain data through the least squares method. 

\section{WAVEFORM DESIGN FOR ACOUSTIC POWER TRANSFER} \label{sec:WAVEFORM DESIGN FOR APT}
	\subsection{Problem Formulation}
	This section firstly investigates a simplified  acoustic power transfer (APT) scenario, where the information-transfer symbol vector $\mathbf{x}_{I}$ is zero and the splitting factor $\rho$ is equal to 1. 
	
	In this case, based on \eqref{eq:zDC_matrix_real}, the harvested power $z_{DC}$ can be rewritten as
	\begin{equation}
		\begin{aligned}
		\label{eq:zDC_APT_matrix}
			z_{\mathrm{DC}}
			&= k_{2}R_{\text{trans}} \mathbf{\hat{w}}_{P}^{T}\mathbf{N}_{0}\mathbf{\hat{w}}_{P}
			+ \frac{3}{2}k_{4}R_{\text{trans}}^{2}\left(\mathbf{\hat{w}}_{P}^{T}\mathbf{N}_{0}\mathbf{\hat{w}}_{P}\right)^{2}\\
			&+ 3k_{4}R_{\text{trans}}^{2}\sum_{k=1}^{N-1}\left(\mathbf{\hat{w}}_{P}^{T}\mathbf{N}_{k}\mathbf{\hat{w}}_{P}\right)^{2}.
			\end{aligned}
	\end{equation}
	
	The optimization problem aims for maximizing the harvested power, which is formulated as
	\begin{subequations} \label{eq:P1}
		\begin{align}
			\text{(P1)}:\quad 
			 \max_{\hat{\mathbf{w}}_{P}} ~~ &z_{\mathrm{DC}}(\hat{\mathbf{w}}_{P}) \tag{\theequation}\label{eq:P1_obj}\\
			\text{s.t.}\quad 
			& ||\hat{\mathbf{w}}_{P}||_{2}^{2} \leq P_{\text{avg}}, \tag{\theequation a}\label{eq:P1a}\\
			& 2\!\left( \sum_{n=1}^{N} \sqrt{\hat{w}_{n,P}^{2} + \hat{w}_{n+N,P}^{2}} \right)^{2} \leq P_{\text{peak}}, \tag{\theequation b}\label{eq:P1b}
		\end{align}
	\end{subequations}
	where $P_{\text{avg}}$ denotes the average transmit power constraint at the transmitter, 
	and $P_{\text{peak}}$ denotes the peak transmit power constraint, which is set to 106~W 
	according to the BII-7511 datasheet \cite{BII7511TransducerDatasheet2025}.
	
	\subsection{Waveform Design}
	Since problem (P1) is non-convex, a first-order Taylor expansion of the 
	objective function $z_{\mathrm{DC}}$ is carried out at the given initial point 
	$\hat{\mathbf{w}}_{P0}$, whereby the non-convex problem is transformed into 
	a convex one. The resulting expression of $z_{\mathrm{DC}}$ is given by
	\begin{equation}
		\tilde{z}(\hat{\mathbf{w}}_{P};\hat{\mathbf{w}}_{P0}) = z_{\mathrm{DC}}(\hat{\mathbf{w}}_{P0})+ \nabla z_{\mathrm{DC}}(\hat{\mathbf{w}}_{P0})^{T} (\hat{\mathbf{w}}_{P}-\hat{\mathbf{w}}_{P0}).
	\end{equation}
	where $\nabla z_{\mathrm{DC}}(\hat{\mathbf{w}}_{P0})$ denotes the gradient of 
	$z_{\mathrm{DC}}$ evaluated at $\hat{\mathbf{w}}_{P0}$.
	
	Consequently, problem (P1) can be reformulated as
	\begin{subequations} \label{eq:P2}
		\begin{align}
			\text{(P2)}:\quad 
			\max_{\hat{\mathbf{w}}_{P}} ~~ &\tilde{z}_{\mathrm{DC}}(\hat{\mathbf{w}}_{P};\hat{\mathbf{w}}_{P0}) \tag{\theequation}\label{eq:P2_obj}\\
			\text{s.t.}\quad 
			& ||\hat{\mathbf{w}}_{P}||_{2}^{2} \leq P_{\text{avg}}, \tag{\theequation a}\label{eq:P2a}\\
			& 2\!\left( \sum_{n=1}^{N} \sqrt{\hat{w}_{n,P}^{2} + \hat{w}_{n+N,P}^{2}} \right)^{2} \leq P_{\text{peak}}. \tag{\theequation b}\label{eq:P2b}
		\end{align}
	\end{subequations}
	
	Problem (P2) can be solved by means of the CVX toolbox in MATLAB~\cite{grant2014cvx}. 
	Since problem (P2) serves as a linear approximation of problem (P1), an iterative scheme 
	is adopted, whereby problem (P2) is repeatedly solved to improve the solution accuracy. The initial Taylor expansion point $\hat{\mathbf{w}}_{P0}$ is randomly generated subject to the feasibility constraints, and in each subsequent iteration the expansion point is updated to the optimal solution obtained in the previous step. This procedure follows the principle of successive convex approximation (SCA), in which the original non-convex problem is iteratively approximated by a sequence of convex subproblems.

	In addition, to enhance the global search capability and mitigate the risk of convergence 
	to local optima, a multiple-initial-points strategy is employed. Specifically, the iterative 
	optimization is executed from several randomly generated initial symbol vectors, and the 
	solution that yields the maximum $z_{\mathrm{DC}}$ is selected as the final result. The detailed 
	procedure is summarized in Algorithm~\ref{alg:multi_carrier_wpt_amp_phase_opt}.
	
	We now evaluate the computational complexity of the proposed algorithm. 
	Let $M_{\text{APT}}$ denote the number of initial points and $I_{\text{APT}}$ 
	denote the number of iterations required by the SCA procedure to solve Problem (P2). Problem (P2), which can be formulated as a second-order cone programming (SOCP) problem, is solved via the interior-point method implemented in CVX. According to~\cite{i.p’olikInteriorPointMethods2010}, 
	the complexity of solving SOCP can be expressed as
	$
		\mathcal{O}\!\left(\sqrt{k}\,\log \frac{1}{\tau}
		\left(m^{3}+ m^{2}n + \sum_{i=1}^{k}n_{i}^{2}\right)\right)
	$ where $k$ is the number of cones, $\tau$ is the solution accuracy, 
	$m$ is the number of equality constraints, $n$ is the total dimension 
	of the second-order cone variables, and $n_i$ denotes the dimension 
	of the $i$-th cone. (Note: all quantities above are derived under the SOCP 
	standard form.) Accordingly, the approximate complexity of solving Problem (P2) can be written as
	$
		\mathcal{O}\!\left( \log\frac{1}{\tau}N^{3.5}\right)
	$. Therefore, the overall computational complexity of 
	Algorithm~\ref{alg:multi_carrier_wpt_amp_phase_opt} is given by
	$
		\mathcal{O}\!\left( M_{\text{APT}}I_{\text{APT}}N^{3.5}\log\frac{1}{\tau}\right)
	$, where $N$ denotes the number of subcarriers.

	\begin{algorithm}[hb!]
		\caption{Algorithm for Optimization in APT System}
		\label{alg:multi_carrier_wpt_amp_phase_opt}
		\begin{algorithmic}[1]
			\REQUIRE System parameters and constraints; required accuracy $\epsilon$; number of initial points $M_{\text{APT}}$  
			\ENSURE Optimal waveform vector $\hat{\mathbf{w}}_{P,\text{opt}}$; optimal $z_{\mathrm{DC},\text{opt}}$
			\STATE $z_{\mathrm{DC},\text{opt}} \gets -\infty$
			\FOR{$k = 1$ \TO $M_{\text{APT}}$}
			\STATE Randomly generate $\hat{\mathbf{w}}_{P}^{(0)}$ satisfying the constraints
			\STATE $z_{\mathrm{DC}}^{(0)} \gets z_{\mathrm{DC}}(\hat{\mathbf{w}}_{P}^{(0)})$
			\STATE $i \gets 0$
			\REPEAT
			\STATE $i \gets i + 1$
			\STATE $\hat{\mathbf{w}}_{P,0} \gets \hat{\mathbf{w}}_{P}^{(i-1)}$
			\STATE $\hat{\mathbf{w}}_{P}^{(i)} \gets \text{solve(P2)}$
			\STATE $z_{\mathrm{DC}}^{(i)} \gets z_{\mathrm{DC}}(\hat{\mathbf{w}}_{P}^{(i)})$
			\UNTIL $\frac{|z_{\mathrm{DC}}^{(i)} - z_{\mathrm{DC}}^{(i-1)}|}{z_{\mathrm{DC}}^{(i-1)}} < \epsilon$
			\IF{$z_{\mathrm{DC}}^{(i)} > z_{\mathrm{DC},\text{opt}}$}
			\STATE $z_{\mathrm{DC},\text{opt}} \gets z_{\mathrm{DC}}^{(i)}$
			\STATE $\hat{\mathbf{w}}_{P,\text{opt}} \gets \hat{\mathbf{w}}_{P}^{(i)}$
			\ENDIF
			\ENDFOR
			\end{algorithmic}
		\end{algorithm}
	
\section{WAVEFORM DESIGN FOR SIMULTANEOUS ACOUSTIC INFORMATION AND POWER TRANSFER} \label{sec:WAVEFORM DESIGN FOR SAIPT}
\subsection{Problem Formulation}
	In the SAIPT system, the objective is to maximize the harvested power 
	by guaranteeing a minimum communication rate requirement. 
	Accordingly, the optimization problem for the SAIPT system is formulated as
		\begin{subequations} \label{eq:P3}
		\begin{align}
			\text{(P3)}:\quad 
			\max_{\rho,\hat{\mathbf{w}}_{I},\hat{\mathbf{w}}_{P}} ~~ &z_{\mathrm{DC}}(\rho,\hat{\mathbf{w}}_{I},\hat{\mathbf{w}}_{P}) \tag{\theequation}\label{eq:P3_obj}\\
			\text{s.t.}\quad 
			& 0<\rho<1 \tag{\theequation a}\label{eq:P3_rho},\\
			& ||\hat{\mathbf{w}}_{I}||_{2}^{2}+||\hat{\mathbf{w}}_{P}||_{2}^{2} \leq P_{\text{avg}}, \tag{\theequation b} \label{eq:P3_Pavg}\\
			& 2\!\left( \sum_{n=1}^{N} \sqrt{\hat{w}_{n,I}^{2} + \hat{w}_{n+N,I}^{2}} + \sqrt{\hat{w}_{n,P}^{2} + \hat{w}_{n+N,P}^{2}} \right)^{2} \notag \\
			&\leq P_{\text{peak}}, \tag{\theequation c}\label{eq:P3_Ppeak}\\
			& R_I \geq R_{\text{th}} \tag{\theequation d}\label{eq:P3_R}.
		\end{align}
	\end{subequations}
	where $R_{\text{th}}$ denotes the threshold of the data rate.
	
	\added{Since problem (P3) involves the joint optimization over the splitting factor $\rho$ and the waveform coefficient vectors $\hat{\mathbf{w}}_{I}$ and $\hat{\mathbf{w}}_{P}$, it leads to a highly non-convex formulation due to the multiplicative coupling among these variables in both the achievable rate constraint and the nonlinear harvested power expression. As a result, the problem does not admit a tractable convex reformulation and is difficult to solve directly using standard optimization techniques.}
	\added{
	To address this challenge, we adopt an alternating optimization (AO) strategy. Specifically, when one group of variables is fixed, the resulting subproblem with respect to the remaining variables exhibits a more structured form and can be efficiently handled using the proposed SCA-based method or a bisection search procedure. Therefore, we first fix the splitting factor $\rho$ and optimize the coefficient vectors $\hat{\mathbf{w}}_{I}$ and $\hat{\mathbf{w}}_{P}$ for information and power transfer, respectively. Then, with $\hat{\mathbf{w}}_{I}$ and $\hat{\mathbf{w}}_{P}$ held fixed, we update $\rho$. These two steps are executed alternately until convergence is achieved.}

\subsection{Transmit Waveform Optimization}
	We first optimize the coefficient vector $\hat{\mathbf{w}}_{I}$ for information transfer and the coefficient vector $\hat{\mathbf{w}}_{P}$ for power transfer by keeping the splitting factor $\rho$ fixed.
	Similar to the previous step in Section~\ref{sec:WAVEFORM DESIGN FOR APT}, to transform the non-convex problem into a convex one, 
	we perform a first-order Taylor expansion of $z_{\mathrm{DC}}$ with respect to 
	$\hat{\mathbf{w}}_{I}$ and $\hat{\mathbf{w}}_{P}$. Accordingly, the objective function 
	$z_{\mathrm{DC}}$ can be approximated as 
	\begin{equation}
		\begin{aligned}
		\tilde{z}_{\mathrm{DC}}(\hat{\mathbf{w}}_{I},\hat{\mathbf{w}}_{P};
		\hat{\mathbf{w}}_{I0},\hat{\mathbf{w}}_{P0}) 
		&= z_{\mathrm{DC}}(\hat{\mathbf{w}}_{I0},\hat{\mathbf{w}}_{P0})\\
		&+ \nabla z_{\mathrm{DC}}(\hat{\mathbf{w}}_{I0},\hat{\mathbf{w}}_{P0})^{T} 
		\begin{pmatrix}
			\hat{\mathbf{w}}_{I}-\hat{\mathbf{w}}_{I0} \\
			\hat{\mathbf{w}}_{P}-\hat{\mathbf{w}}_{P0}
		\end{pmatrix},
		\end{aligned}
	\end{equation}
		where $\nabla z_{\mathrm{DC}}(\hat{\mathbf{w}}_{I0},\hat{\mathbf{w}}_{P0})$ denotes the gradient of 
	$z_{\mathrm{DC}}$ evaluated at $(\hat{\mathbf{w}}_{I0}, \hat{\mathbf{w}}_{P0})$.
	
	Hence, problem (P3) can be transformed into
	\begin{subequations} \label{eq:P4}
		\begin{align}
			\text{(P4)}:
			\max_{\hat{\mathbf{w}}_{I},\hat{\mathbf{w}}_{P}} ~~ &\tilde{z}_{\mathrm{DC}}(\hat{\mathbf{w}}_{I},\hat{\mathbf{w}}_{P};\hat{\mathbf{w}}_{I0},\hat{\mathbf{w}}_{P0}) \tag{\theequation}\label{eq:P4_obj}\\
			\text{s.t.}\quad 
			& ||\hat{\mathbf{w}}_{P}||_{2}^{2}+||\hat{\mathbf{w}}_{I}||_{2}^{2} \leq P_{\text{avg}}, \tag{\theequation a} \label{eq:P4_Pavg}\\
			& 2\!\left( \sum_{n=1}^{N} \sqrt{\hat{w}_{n,P}^{2} + \hat{w}_{n+N,P}^{2}} + \sqrt{\hat{w}_{n,I}^{2} + \hat{w}_{n+N,I}^{2}} \right)^{2} \leq P_{\text{peak}}, \tag{\theequation b}\label{eq:P4_Ppeak}\\
			&R_I  \geq R_{\text{th}} \tag{\theequation c}\label{eq:P4_R}.
		\end{align}
	\end{subequations}
	
	\added{To handle the non-convexity of constraint~\eqref{eq:P4_R}, auxiliary variables $\{s_n\}$ and $\{t_n\}$ are introduced as
	\begin{equation}
	\begin{aligned}
	e^{s_{n}}  =&(1-\rho)\sigma_{n}^{2}+\sigma_{cov}^{2}+(1-\rho)\eta_{n}\hat{h}_{n}^{2}\sigma_{e}^{2}\left(\hat{w}^{2}_{n,P}+\hat{w}^{2}_{n+N,P}\right)\\
	&+(1-\rho)\eta_{n}(1+\sigma_{e}^{2})\hat{h}_{n}^{2}\left(\hat{w}^{2}_{n,I}+\hat{w}^{2}_{n+N,I}\right) \\
	e^{t_{n}}  =&(1-\rho)\sigma_{n}^{2}+\sigma_{cov}^{2}+(1-\rho)\eta_{n}\hat{h}_{n}^{2}\sigma_{e}^{2}\left(\hat{w}^{2}_{n,P}+\hat{w}^{2}_{n+N,P}\right)
	\end{aligned}, \forall n,
	\end{equation}}
	\added{Therefore, problem (P4) can be further reformulated as}
	{\color{added}
	\begin{subequations} \label{eq:P5}
			\begin{align}
			&\text{(P5)}:\quad 
			\max_{\hat{\mathbf{w}}_{I},\hat{\mathbf{w}}_{P},\{s_n\}} ~~ \tilde{z}_{\mathrm{DC}}(\hat{\mathbf{w}}_{I},\hat{\mathbf{w}}_{P};\hat{\mathbf{w}}_{I0},\hat{\mathbf{w}}_{P0}) \tag{\theequation}\label{eq:P5_obj}\\
			\text{s.t.}\quad 
			& ||\hat{\mathbf{w}}_{P}||_{2}^{2}+||\hat{\mathbf{w}}_{I}||_{2}^{2} \leq P_{\text{avg}}, \tag{\theequation a} \label{eq:P5_Pavg}\\
			& 2\!\left( \sum_{n=1}^{N} \sqrt{\hat{w}_{n,P}^{2} + \hat{w}_{n+N,P}^{2}} + \sqrt{\hat{w}_{n,I}^{2} + \hat{w}_{n+N,I}^{2}} \right)^{2} \leq P_{\text{peak}}, \tag{\theequation b}\label{eq:P5_Ppeak}\\
			& \Delta f\sum_{n=1}^{N}(s_{n}-t_n)\log_{2}(e) \geq R_{\text{th}},\tag{\theequation c}\label{eq:P5_R}\\
			& (1-\rho)\sigma_{n}^{2}+\sigma_{cov}^{2}+(1-\rho)\eta_{n}\hat{h}_{n}^{2}\sigma_{e}^{2}\left(\hat{w}^{2}_{n,P}+\hat{w}^{2}_{n+N,P}\right) \notag \\
			&~ +(1-\rho)\eta_{n}(1+\sigma_{e}^{2})\hat{h}_{n}^{2}\left(\hat{w}^{2}_{n,I}+\hat{w}^{2}_{n+N,I}\right) \geq e^{s_n},\forall n, \tag{\theequation d} \label{eq:P5_s}\\
			&(1-\rho)\sigma_{n}^{2}+\sigma_{cov}^{2}
			+(1-\rho)\eta_{n}\hat{h}_{n}^{2}\sigma_{e}^{2}\left(\hat{w}^{2}_{n,P}+\hat{w}^{2}_{n+N,P}\right) \leq e^{t_n},\forall n. \tag{\theequation e} \label{eq:P5_t}
		\end{align}
	\end{subequations}}
	
	\added{By further performing a first-order Taylor expansion of \eqref{eq:P5_s} and \eqref{eq:P5_t}, problem (P5) can be reformulated as}
		{\color{added}
		\interdisplaylinepenalty=2500
		\begin{subequations} \label{eq:P6}
		\begin{align}
			&\text{(P6)}:\quad 
			\max_{\hat{\mathbf{w}}_{I},\hat{\mathbf{w}}_{P},\{s_n\}} ~~ \tilde{z}_{\mathrm{DC}}(\hat{\mathbf{w}}_{I},\hat{\mathbf{w}}_{P};\hat{\mathbf{w}}_{I0},\hat{\mathbf{w}}_{P0}) \tag{\theequation}\label{eq:P6_obj}\\
			\text{s.t.}\quad 
			& ||\hat{\mathbf{w}}_{P}||_{2}^{2}+||\hat{\mathbf{w}}_{I}||_{2}^{2} \leq P_{\text{avg}}, \tag{\theequation a} \label{eq:P6_Pavg}\\
			 2&\!\left( \sum_{n=1}^{N} \sqrt{\hat{w}_{n,P}^{2} + \hat{w}_{n+N,P}^{2}} + \sqrt{\hat{w}_{n,I}^{2} + \hat{w}_{n+N,I}^{2}} \right)^{2}\leq P_{\text{peak}}, \tag{\theequation b}\label{eq:P6_Ppeak}\\
			& \Delta f\sum_{n=1}^{N}(s_{n}-t_n)\log_{2}(e) \geq R_{\text{th}},\tag{\theequation c}\label{eq:P6_R}\\
			&(1-\rho)\sigma_{n}^{2}+\sigma_{cov}^{2}  +(1-\rho)\eta_{n}\hat{h}_{n}^{2}\sigma_{e}^{2}\Bigg(2\hat{w}^{(0)}_{n,P}\hat{w}_{n,P}-\left( \hat{w}^{(0)}_{n,P}\right)^{2} \notag
			\\
			&\quad+2\hat{w}^{(0)}_{n+N,P}\hat{w}_{n+N,P}-\left( \hat{w}^{(0)}_{n+N,P}\right)^{2}\Bigg) \notag \\ 
			&\quad+(1-\rho)\eta_{n}(1+\sigma_{e}^{2})\hat{h}_{n}^{2}\Bigg(2\hat{w}^{(0)}_{n,I}\hat{w}_{n,I}-\left( \hat{w}^{(0)}_{n,I}\right)^{2} \notag \\ 
			&\quad+2\hat{w}^{(0)}_{n+N,I}\hat{w}_{n+N,I}-\left( \hat{w}^{(0)}_{n+N,I}\right)^{2}\Bigg) \geq e^{s_{n}},~\forall n, \tag{\theequation d} \label{eq:P6_s}\\
			&(1-\rho)\sigma_{n}^{2}+\sigma_{cov}^{2}+(1-\rho)\eta_{n}\hat{h}_{n}^{2}\sigma_{e}^{2}\left(\hat{w}^{2}_{n,P}+\hat{w}^{2}_{n+N,P}\right)\notag   \\ 
			&\leq e^{t^{(0)}_{n}}+e^{t^{(0)}_{n}}(t_{n}-t_{n}^{(0)}),~\forall n,\tag{\theequation e}
		\end{align}
	\end{subequations}}
	\added{where $\hat{w}_{n,I}^{(0)}$ and  $\hat{w}_{n,P}^{(0)}$ denote the $n$-th component of $\hat{\mathbf{w}}_{I0}$ and  $\hat{\mathbf{w}}_{P0}$, respectively.}
	
	\added{Consequently, by applying the rate transformation to the rate constraint and performing first-order Taylor expansions of both the objective function and the constraint, the waveform vectors can be optimized via the successive convex approximation (SCA) method by iteratively solving problem (P6). At each iteration, the first-order Taylor expansion points $\hat{\mathbf{w}}_{I0}$ and $\hat{\mathbf{w}}_{P0}$ are updated to the waveform vectors obtained from the previous iteration, whereas the initial expansion points are randomly generated feasible vectors that satisfy the constraints.}
	
\subsection{Power Splitting Factor Adjustment}
	This section focuses on optimizing the splitting factor $\rho$ with the 
	waveform vectors $\hat{\mathbf{w}}_{I}$ and $\hat{\mathbf{w}}_{P}$ held fixed. 
	The optimization problem can be expressed as
	\begin{subequations} \label{eq:P7}
		\begin{align}
			\text{(P7)}:\quad 
			&\max_{\rho} ~~ {z}_{\mathrm{DC}}(\rho) \tag{\theequation}\label{eq:P7_obj}\\
			\text{s.t.}\quad &
			 0<\rho<1, \tag{\theequation a} \label{eq:P7_rho}\\
			& R_I \geq R_{\text{th}}. \tag{\theequation b}\label{eq:P7_R}
		\end{align}
	\end{subequations}
	
	Since $z_{\mathrm{DC}}$ increases monotonically with $\rho$, a bisection search can be 
	employed to determine the value of $\rho$ that exactly satisfies the rate 
	constraint, thereby achieving the maximum $z_{\mathrm{DC}}$. The detailed procedure is summarized in Algorithm~\ref{alg:bisection_rho}.
	
	\begin{algorithm}[tbh!]
		\caption{Bisection Search for Optimizing the Splitting Factor $\rho$}
		\label{alg:bisection_rho}
		\begin{algorithmic}[1]
			\REQUIRE Required accuracy $\epsilon_{\text{SAIPT}}$, initial interval $[\rho_{\min}, \rho_{\max}]$, rate threshold $R_{\text{th}}$
			\ENSURE Optimal splitting factor $\rho^{*}$
			\WHILE{$\rho_{\max}-\rho_{\min} > \epsilon_{\text{SAIPT}}$}
			\STATE $\rho \gets (\rho_{\min}+\rho_{\max})/2$
			\IF{$R_I(\rho) \geq R_{\text{th}}$}
			\STATE $\rho_{\min} \gets \rho$ \quad 
			\ELSE
			\STATE $\rho_{\max} \gets \rho$ \quad 
			\ENDIF
			\ENDWHILE
			\STATE $\rho^{*} \gets \rho_{\min}$
		\end{algorithmic}
	\end{algorithm}
	
	Ultimately, by alternately optimizing the waveform vectors $\hat{\mathbf{w}}_{I}$ 
	and $\hat{\mathbf{w}}_{P}$ with a fixed splitting factor $\rho$, and optimizing $\rho$ 
	with fixed waveform vectors, the entire set of variables can be jointly optimized. 
	This alternating optimization (AO) procedure is iterated until convergence, 
	thereby achieving an efficient solution to the original problem.
	
	Similarly, to avoid convergence to a local optimum, the AO procedure is executed 
	from multiple randomly generated initial points. Among the 
	obtained solutions, the one that maximizes the DC output is selected as the 
	optimal result. The detailed procedure is summarized in Algorithm~\ref{alg:AO in SAIPT}.
		\begin{algorithm}[ht!]
		\caption{Algorithm for Alternating Optimization in SAIPT System}
		\label{alg:AO in SAIPT}
		\begin{algorithmic}[1]
			\REQUIRE System parameters and constraints; required accuracy $\epsilon$; number of initial points $M_{\text{SAIPT}}$  
			\ENSURE Optimal waveform vectors $\hat{\mathbf{w}}_{I,\text{opt}}$ and $\hat{\mathbf{w}}_{P,\text{opt}}$; optimal splitting factor $\rho_{\text{opt}}$;\\
			\quad ~~~~optimal $z_{\mathrm{DC},\text{opt}}$
			\STATE $z_{\mathrm{DC},\text{opt}} \gets -\infty$
			\FOR{$k = 1$ \TO $M_{\text{SAIPT}}$}
			\STATE Randomly generate $\rho^{(0)}$, $\hat{\mathbf{w}}_{P}^{(0)}$, $\hat{\mathbf{w}}_{I}^{(0)}$ satisfying the constraints
			\STATE $z_{\mathrm{DC}}^{(0)} \gets z_{\mathrm{DC}}(\rho^{(0)},\hat{\mathbf{w}}_{I}^{(0)},\hat{\mathbf{w}}_{P}^{(0)})$
			\STATE $i \gets 0$
			\REPEAT
			\STATE $i \gets i + 1$
			\STATE $\hat{\mathbf{w}}_{I}^{(i)},\hat{\mathbf{w}}_{P}^{(i)} \gets \text{solve (P6) iteratively until convergence}$, 
			following a procedure similar to that in Algorithm~\ref{alg:multi_carrier_wpt_amp_phase_opt}
			of Section~\ref{sec:WAVEFORM DESIGN FOR APT}.
			\STATE $\rho^{(i)} \gets \text{solve (P7)}$
			\STATE $z_{\mathrm{DC}}^{(i)} \gets z_{\mathrm{DC}}(\rho^{(i)},\hat{\mathbf{w}}_{I}^{(i)},\hat{\mathbf{w}}_{P}^{(i)})$
			\UNTIL $\frac{|z_{\mathrm{DC}}^{(i)} - z_{\mathrm{DC}}^{(i-1)}|}{z_{\mathrm{DC}}^{(i-1)}} < \epsilon$
			\IF{$z_{\mathrm{DC}}^{(i)} > z_{\mathrm{DC},\text{opt}}$}
			\STATE $z_{\mathrm{DC},\text{opt}} \gets z_{\mathrm{DC}}^{(i)}$
			\STATE $\rho_{\text{opt}},\hat{\mathbf{w}}_{I,\text{opt}},\hat{\mathbf{w}}_{P,\text{opt}} \gets \rho^{(i)},\hat{\mathbf{w}}_{I}^{(i)},\hat{\mathbf{w}}_{P}^{(i)}$
			\ENDIF
			\ENDFOR
		\end{algorithmic}
	\end{algorithm}
	
		\begin{figure*}[htbp]
		\begin{minipage}{0.45\textwidth}
			\centering
			\includegraphics[width=\linewidth]{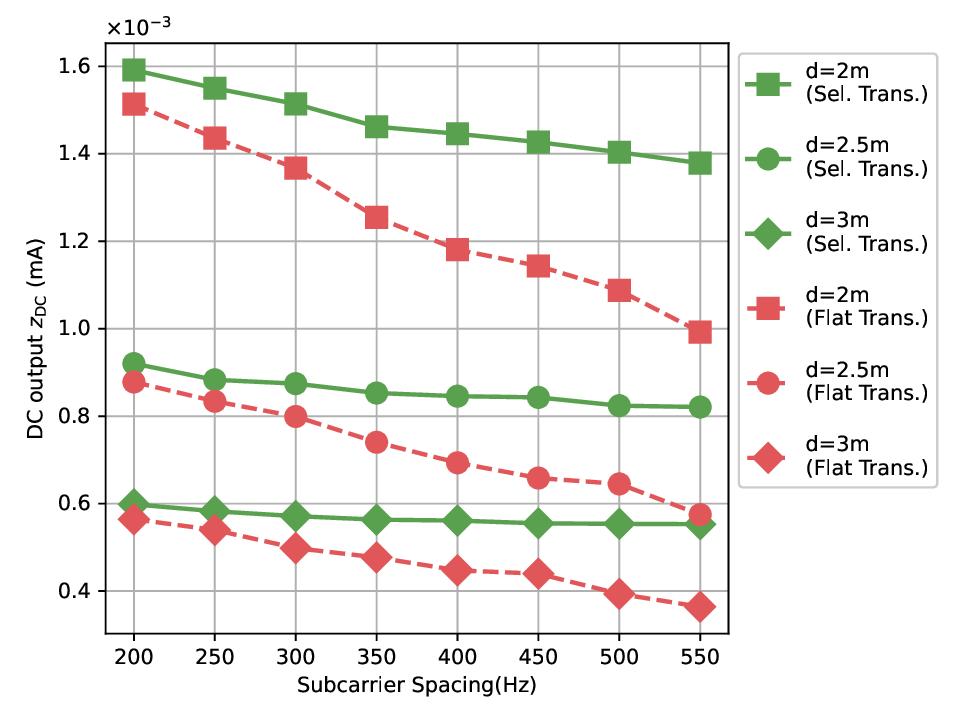}
			\vspace{-25pt}
			\caption{\added{DC output versus subcarrier spacing $\Delta f$ at $N=20$, $P_{avg}=2W$, $R_{\text{th}}=2000bps$ and $\sigma_e^2=0.05$ in SAIPT. Results are shown for both ``Sel. Trans.'' and ``Flat Trans.'' cases.}\protect\\ \phantom{Extra line}\protect\\ \phantom{Extra line}}
			\label{fig:SAIPT_trans_compare}
		\end{minipage}\hfill
		\begin{minipage}{0.45\textwidth}
			\centering
			\includegraphics[width=\linewidth]{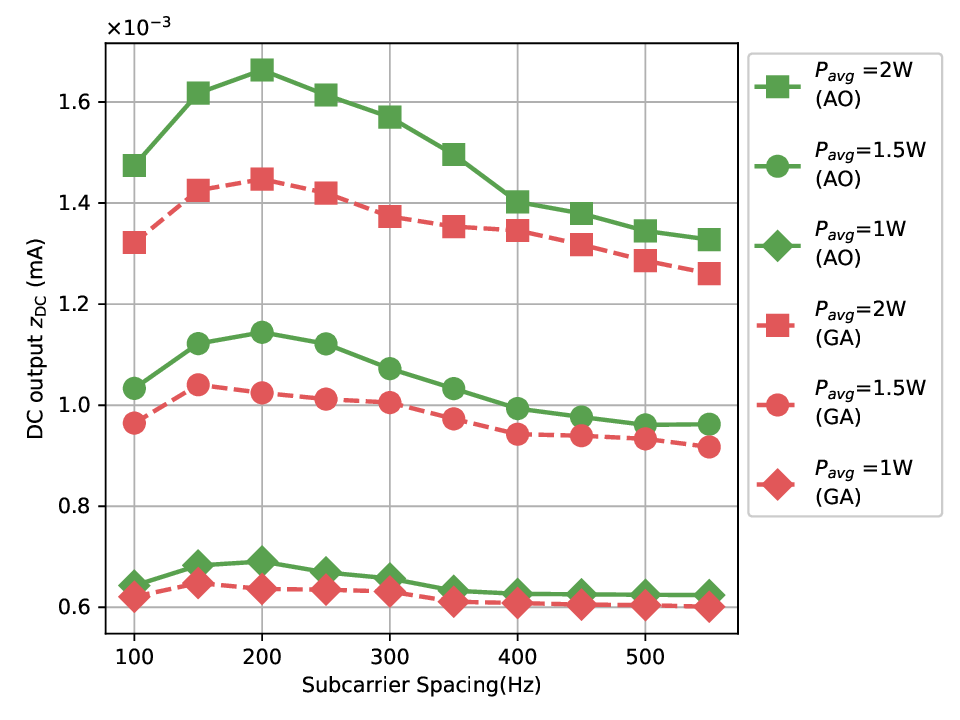}
			\vspace{-25pt}
			\caption{DC output versus subcarrier spacing $\Delta f$ at $N=20$, $d=2m$, $R_{\text{th}}=10Kbps$ and $\sigma_e^2=0.05$ in SAIPT. The label ``AO'' refers to the proposed alternating optimization algorithm,
			whereas ``GA'' denotes the genetic algorithm employed
			as a benchmark scheme. The GA is implemented with a population size of 200 and
			terminated after $10^5$ generations or when convergence is reached.}
			\label{fig:SAIPT_GA_compare}
		\end{minipage}
	\end{figure*} 
	\begin{figure*}[htbp]
		\begin{minipage}{0.45\textwidth}
			\centering
			\includegraphics[width=\linewidth]{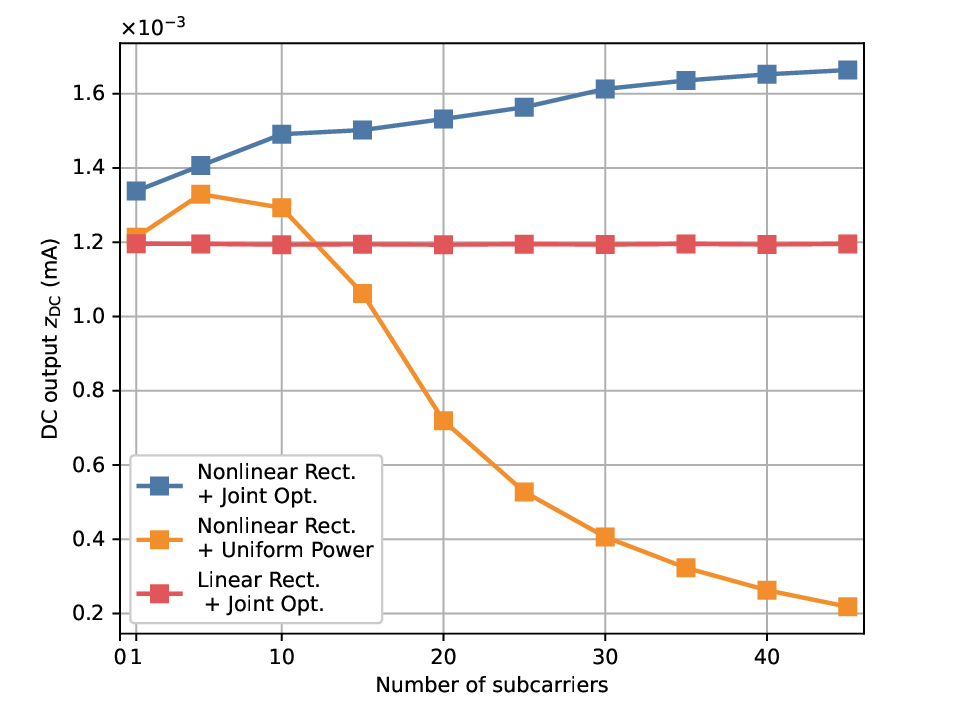}
			\vspace{-25pt}
			\caption{\added{DC output versus number of subcarriers $N$ at $P_{avg}=2$ W, $d=2$ m, $\Delta f=300$ Hz, $R_{\text{th}}=2000$ bps, and $\sigma_e^2=0.01$ in SAIPT. ``Nonlinear Rect.'' denotes the proposed nonlinear rectifier model, while ``Linear Rect.'' refers to the linear rectifier model adopted from \cite{esmaiel2020wireless}. ``Joint Opt.'' represents the proposed joint waveform optimization scheme, whereas ``Uniform Power'' corresponds to equal power allocation for both information and power waveforms across all subcarriers.}}
			\label{fig:SAIPT_N_baseline}
		\end{minipage}\hfill
		\begin{minipage}{0.45\textwidth}
			\centering
			\includegraphics[width=\linewidth]{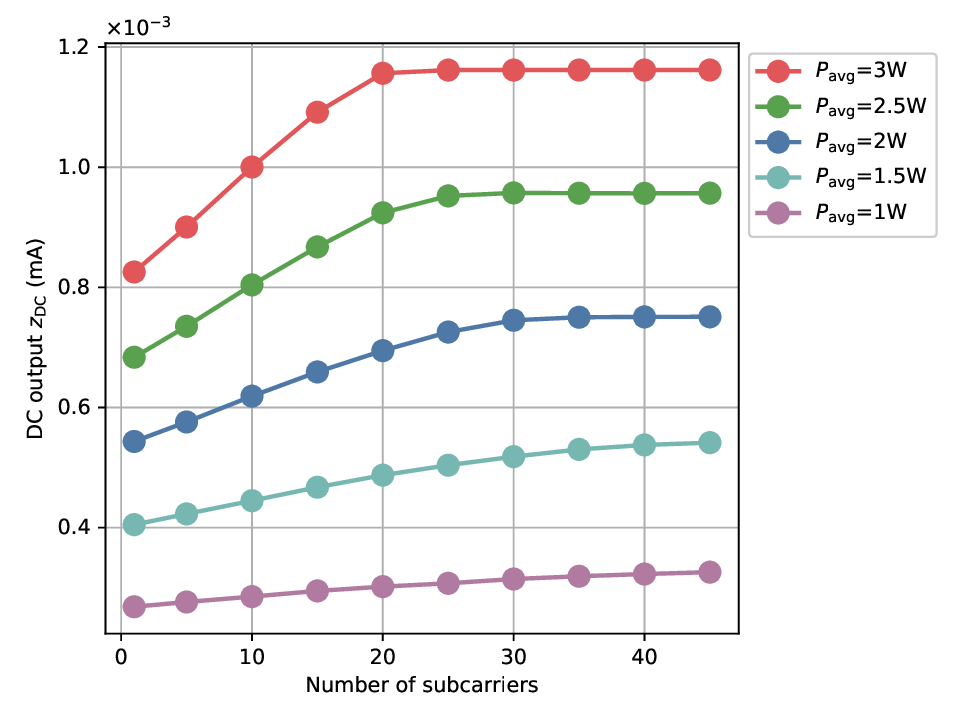}
			\vspace{-25pt}
			\caption{DC output versus number of subcarriers $N$ at $d=3m$ and $\Delta f=100Hz$ in APT.\protect\\ \phantom{Extra line}\protect\\ \phantom{Extra line}\protect\\ \phantom{Extra line}\protect\\ \phantom{Extra line}\protect\\ \phantom{Extra line}}
			\label{fig:APT_N_P}
		\end{minipage}
	\end{figure*} 
	
\subsection{Computational Complexity Analysis}
	We first define $I_{1}$ as the number of iterations required by the SCA procedure to solve Problem (P6), and $I_{2}$ as the number of iterations of the alternating optimization. Similar to Problem (P2), the complexity of solving Problem (P6) can be approximated as $\mathcal{O}_1=
	\mathcal{O}\!\left( \log\frac{1}{\tau}N^{3.5}\right)
	$. Additionally, the complexity of solving Problem (P7) via bisection search is formulated as $\mathcal{O}_{2}=\mathcal{O}\!\left( \log\frac{1}{\epsilon_{\text{SAIPT}}}\right)$. Therefore, the overall computational complexity of alternating optimization can be expressed as $\mathcal{O}_{3}=M_{\text{SAIPT}}I_{1}I_{2}\left( \mathcal{O}_{1}+\mathcal{O}_{2}
	\right)$. Since $\mathcal{O}_{1}\gg \mathcal{O}_{2}$, the total complexity can be further approximated as
	$\mathcal{O}_{3}= \mathcal{O}\!\left( M_{\text{SAIPT}}I_{1}I_{2}N^{3.5}\log\tfrac
	{1}{\tau}\right)$, where $N$ denotes the number of subcarriers.
	
\begin{table}[h]
	\centering
	\caption{Simulation parameters}
	\label{tab:parameter_settings}
	\begin{tabular}{lc}
		\toprule
		\textbf{Parameters} & \textbf{Values} \\
		\midrule
		Resonant frequency of transducer, $f_r$ & $48.3\,\text{kHz}$ \\
		Overall efficiency of the electrical circuitry, $\Phi$ & 0.8 \\
		Down-conversion induced noise, $\sigma_{\text{cov}}^2$ & $-80\,\text{dBm}$ \\
		Number of initial points of APT, $M_{\text{APT}}$&100\\
		Number of initial points of SAIPT, $M_{\text{SAIPT}}$&200\\
		\bottomrule
	\end{tabular}
\end{table}

\section{SIMULATION RESULTS}  \label{sec:SIMULATION RESULTS}
	In this section, simulation results are presented to evaluate the DC output performance of both APT and SAIPT.  For the transducer and rectifier, the parameters are set as $k_{2}=0.0034$, $k_{4}=0.3859$, and $R_{\text{trans}}=50\Omega$. Table~\ref{tab:parameter_settings} lists the other simulation parameters.

	\subsection{Performance Comparison and Discussion}
	Fig.~\ref{fig:SAIPT_trans_compare} illustrates the influence of the transducer response on system performance in SAIPT. The DC output decreases with the enlargement of the subcarrier spacing, since a larger spacing pushes more subcarriers away from the transducer’s resonance frequency, thereby reducing the electro-acoustic and acousto-electric conversion efficiency. Furthermore, the discrepancy between the two cases is further amplified as the spacing increases. \added{The result highlights the importance of incorporating the frequency-selective characteristics of the transducer into performance evaluation, thereby improving the accuracy of system analysis and the effectiveness of subsequent waveform optimization.}

	\added{Fig.~\ref{fig:SAIPT_GA_compare} compares the performance of the proposed
	alternating optimization (AO) algorithm and the genetic algorithm (GA)
	under different transmit power levels in the SAIPT system.
	At the initial stage, increasing $\Delta f$ enlarges the communication bandwidth, allowing more power to be allocated from information transmission to energy transfer, thereby enhancing the DC output. However, as $\Delta f$ further increases, subcarriers move away from the resonant frequency of the transducer, leading to reduced energy conversion efficiency and a consequent decrease in DC output. Moreover, for all considered transmit power levels, the proposed AO algorithm achieves higher DC output than the GA scheme. This advantage arises because the AO algorithm explicitly exploits the mathematical structure of the nonlinear rectifier model and the frequency-dependent transducer response, whereas the GA performs a general-purpose stochastic search without leveraging such problem-specific properties.}
	
	\begin{figure*}[t]
		\centering
		{\subfigure[]{\includegraphics[width=0.4\linewidth]{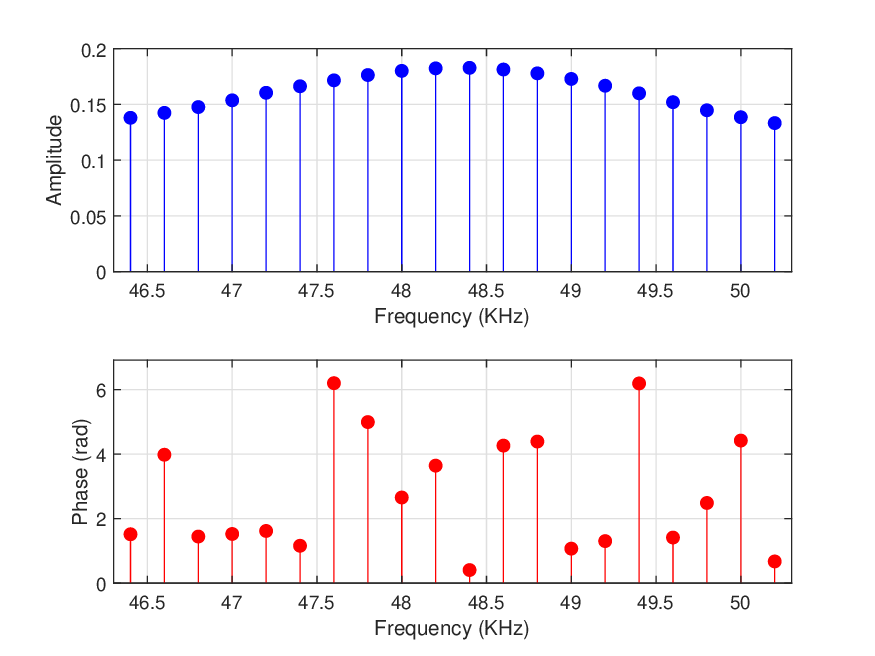}}
			\hfil
			\subfigure[]{\includegraphics[width=0.4\linewidth]{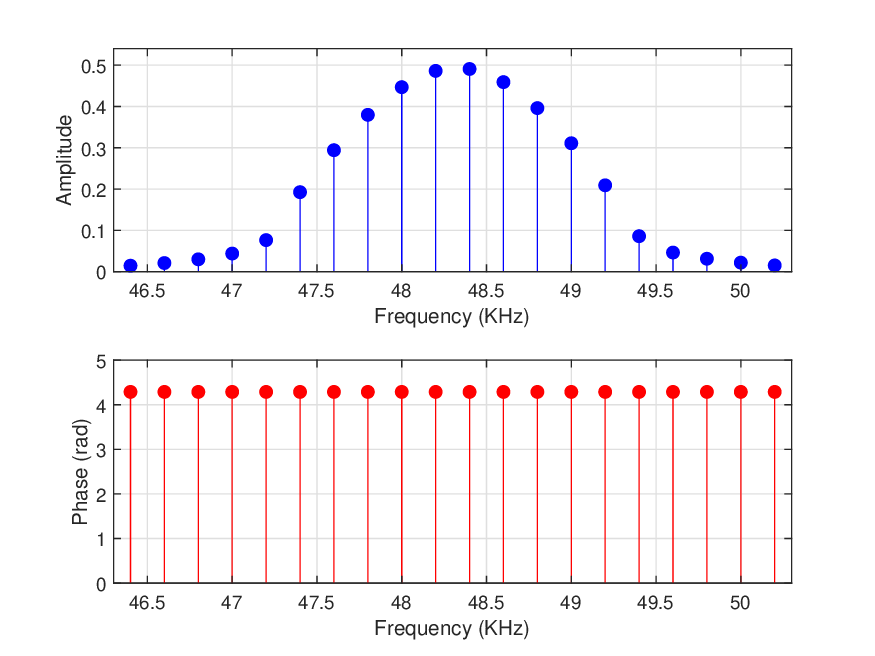}}}
		\setlength{\abovecaptionskip}{0pt}
		\setlength{\belowcaptionskip}{0pt}
		\caption{\added{Waveform shaping results of SAIPT with $N=20$, $P_{avg}=2$ W, $d=2$ m, $\Delta f=200$ Hz, $R=22$ Kbps, and $\sigma_e^2=0.02$: 
				(a) Amplitude and phase of the information transfer vector $\hat{\mathbf{w}}_{I}$; 
				(b) Amplitude and phase of the power transfer vector $\hat{\mathbf{w}}_{P}$.}}
		\label{fig:SAIPT_opt_results}
	\end{figure*} 
	
	\begin{figure*}[tp]
		\begin{minipage}{0.45\textwidth}
			\centering
			\includegraphics[width=\linewidth]{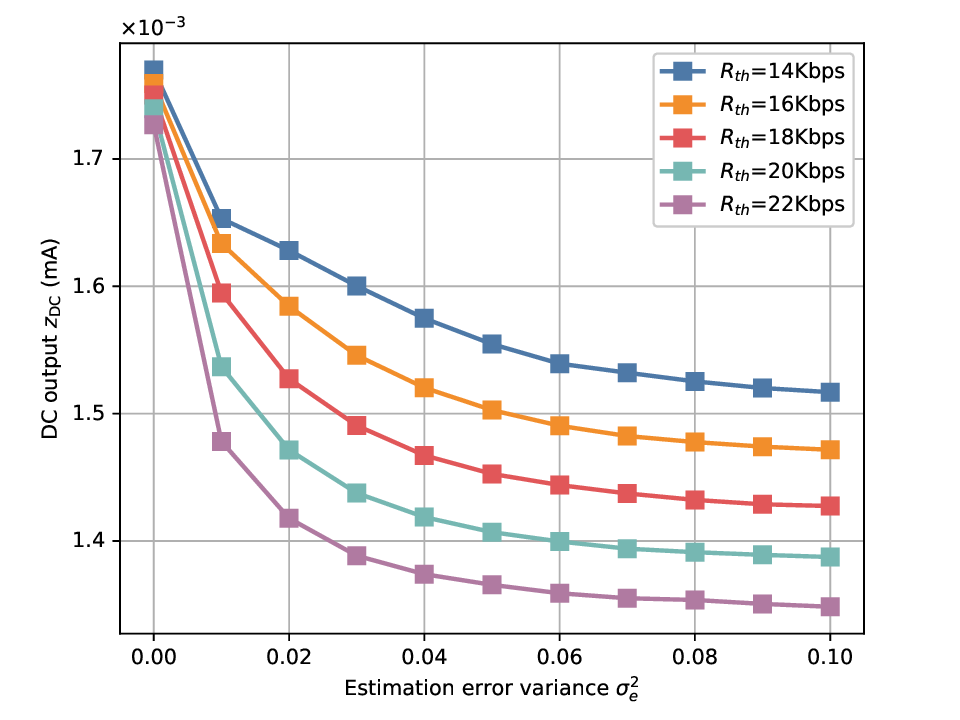}
			\vspace{-25pt}
			\caption{\added{DC output versus estimation error variance  $\sigma_e^2$ at $N=20$, $P_{avg}=2W$, $d=2m$ and $\Delta f=200Hz$ in SAIPT.}}
			\label{fig:SAIPT_err_R}
		\end{minipage}\hfill
		\begin{minipage}{0.45\textwidth}
			\centering
			\includegraphics[width=\linewidth]{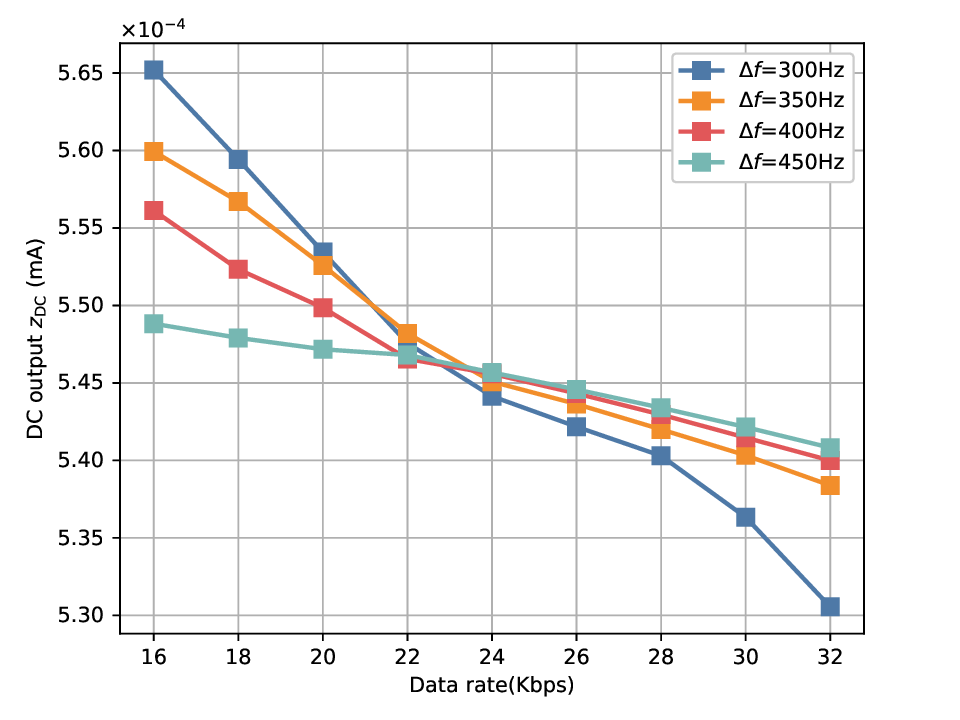}
			\vspace{-25pt}
			\caption{\added{DC output versus data rate $R_{\text{th}}$ at $N=20$, $P_{\text{avg}}=2W$ , $d=3m$ and $\sigma_e^2=0.05$ in SAIPT.}}
			\label{fig:SAIPT_R_Deltaf}
		\end{minipage}
	\end{figure*} 
	
	\added{Fig.~\ref{fig:SAIPT_N_baseline} depicts the DC output versus the number of subcarriers $N$ under different rectifier models and waveform design strategies. With the nonlinear rectifier model and joint optimization, the DC output increases monotonically with $N$, demonstrating the advantage of multi-carrier transmission. As $N$ grows, the resulting waveform exhibits a higher PAPR, which enhances the higher-order components under the nonlinear rectifier and thus improves the harvested DC output. Under the nonlinear rectifier model with uniform power allocation, the
	DC output first increases and then decreases as $N$ grows. The initial gain results from the PAPR enhancement brought by multi-carrier transmission, while further increasing $N$ allocates more power to frequency components with lower transducer conversion efficiency, leading to performance degradation. For the linear rectifier model adopted from \cite{esmaiel2020wireless} with joint optimization, the DC output remains nearly constant with respect to $N$. Since the linear model neglects the higher-order nonlinear terms of the rectifier, the harvested DC power depends solely on the average received power, rendering it insensitive to the PAPR variation induced by multi-carrier transmission. Moreover, without the contribution of higher-order components, the resulting DC output is lower than that achieved under the nonlinear rectifier model. The result highlights that incorporating the nonlinear rectifier characteristics enables effective exploitation of the high-PAPR property of multi-carrier waveforms to enhance the harvested DC output, and also underscores the importance of adapting the waveform design to the transducer’s frequency selectivity for performance improvement.}
		
	\subsection{Evaluation of APT}
	\added{
	In the simplified APT scenario, Fig.~\ref{fig:APT_N_P} shows the variation of the DC output $z_{\text{DC}}$ with respect to the number of subcarriers $N$ under different transmit power levels $P_{\text{avg}}$. It is observed that $z_{\text{DC}}$ increases superlinearly with $P_{\text{avg}}$, owing to the higher-order nonlinear terms in the rectifier model, which become more dominant under stronger excitation. As $N$ increases, $z_{\text{DC}}$ first rises due to the higher PAPR of multicarrier signals, which enhances the nonlinear rectification effect. However, when $N$ becomes sufficiently large, the growth gradually slows down and tends to saturate. The saturation behavior can be attributed to the limited effective bandwidth of the transducer, as additional subcarriers are allocated to frequency regions with low conversion efficiency, while the instantaneous peak power is further constrained by the transducer characteristics. These results indicate that increasing $N$ does not always bring proportional gains in harvested energy.  Instead, an appropriate choice of $N$ is required to balance PAPR enhancement, transducer frequency response, and peak-power limitations in practical APT system design.}

	\subsection{Evaluation of SAIPT}
	The optimized waveform vectors under a given set of system parameters are shown in Fig.~\ref{fig:SAIPT_opt_results}, where the optimized power splitting factor is $\rho = 0.96$. In terms of amplitude, both the information transfer vector $\hat{\mathbf{w}}_{I}$ and the power transfer vector $\hat{\mathbf{w}}_{P}$ exhibit larger amplitudes at central subcarriers while decreasing towards the edges, indicating that more power is allocated to the subcarriers close to the transducer's resonant frequency to enhance energy conversion efficiency, while simultaneously leveraging the high PAPR of multicarrier waveforms to improve the rectifier’s efficiency. In terms of phase, since the information transfer symbols $\{x_{n,I}\}$ are random, the phases of $\hat{\mathbf{w}}_{I}$ do not follow any clear pattern; in contrast, the power transfer vector $\hat{\mathbf{w}}_{P}$ corresponds to a deterministic signal and therefore exhibits phase consistency.

	\added{
	Fig.~\ref{fig:SAIPT_err_R} illustrates the DC output $z_{\mathrm{DC}}$ versus the channel estimation error variance $\sigma_e^2$ under different rate requirements $R_{\mathrm{th}}$. The DC output decreases as the rate requirement increases, since a larger portion of the transmit power must be allocated to information transfer in order to satisfy the rate constraint, leaving less power available for energy transfer.
    In addition, as $\sigma_e^2$ increases, imperfect CSI leads to stronger residual interference from the power-transfer symbols to the information symbols. This reduces the effective SINR and makes the rate constraint more difficult to satisfy. As a result, more power is allocated to information transfer, which reduces the harvested DC output. This effect becomes more pronounced under higher rate requirements, resulting in an increasingly larger performance gap between different rate constraints. This result highlights the sensitivity of energy transfer performance to channel estimation accuracy, especially under stringent rate requirements.
	}
	
	\added{	The effect of subcarrier spacing on the relationship between $z_{\text{DC}}$ and $R_{\text{th}}$ is shown in Fig.~\ref{fig:SAIPT_R_Deltaf}. At the initial stage, a smaller subcarrier spacing concentrates the subcarriers around the resonant frequency of the transducer, leading to higher energy conversion efficiency and greater DC output. However, as the data rate increases, the limited bandwidth associated with small subcarrier spacing results in more power being allocated to information transfer, thereby yielding a lower DC output compared with larger subcarrier spacing. This result highlights the necessity of properly selecting the subcarrier spacing to achieve a trade-off between energy conversion efficiency and communication bandwidth.}

\section{CONCLUSION} \label{sec:CONCLUSION}
	This paper presented a study on the SAIPT system that takes into account the characteristics of the transducer and the nonlinear behavior of the rectifier circuit, where the waveform vectors are optimized in APT and jointly optimized with the power splitting factor in SAIPT to maximize the DC output, subject to constraints on transmit power, peak power, and achievable data rate. Simulation results confirm the effectiveness and necessity of incorporating both the transducer characteristics and the rectifier nonlinearity into the performance evaluation. They also show that appropriate selection of system parameters, such as the number of subcarriers and the subcarrier spacing, can fully exploit the characteristics of the transducer, the rectifier nonlinearity, and the PAPR of multicarrier waveforms to significantly enhance the DC output. Additionally, the results highlight the trade-offs between information and power transfer, as well as between the transducer and rectifier characteristics.

\bibliographystyle{IEEEtran}
\bibliography{IEEEabrv,reference}
\end{document}